\documentclass[twocolumn,twocolappendix]{openjournal}
\renewcommand{\baselinestretch}{1.075} 
\shortauthors{Mendoza et al.}
\graphicspath{{./}{figures/}}
\usepackage{macros}

\usepackage[most]{tcolorbox}
\newtcolorbox{jeffbox}{
  colback=blue!5,       
  colframe=red,         
  coltext=red,          
  sharp corners,
  boxrule=0.8pt,
  left=4pt,
  right=4pt,
  top=2pt,
  bottom=2pt,
  fonttitle=\bfseries,
  before skip=5pt,
  after skip=5pt,
}

\begin{document}

\title{Simulation-Based Inference for Probabilistic Galaxy Detection and Deblending}

\author{\href{https://orcid.org/0000-0002-6313-4597}{Ismael Mendoza}}
\affiliation{Department of Astronomy, University of Maryland, College Park, MD 20742}
\affiliation{Department of Physics, University of Michigan, Ann Arbor, MI 48109}
\affiliation{Leinweber Institute for Theoretical Physics, University of Michigan, Ann Arbor, MI 48109, USA}

\author{Derek Hansen}
\affiliation{Department of Statistics, University of Michigan, Ann Arbor, MI 48109}

\author{Runjing Liu}
\affiliation{The Voleon Group, Berkeley, CA 94704}

\author{Zhe Zhao}
\affiliation{Department of Statistics, University of Michigan, Ann Arbor, MI 48109}

\author{Ziteng Pang}
\affiliation{Department of Statistics, University of Michigan, Ann Arbor, MI 48109}

\author{\href{https://orcid.org/0000-0002-5068-7918}{Axel Guinot}}
\affiliation{Department of Physics, McWilliams Center for Cosmology and Astrophysics, Carnegie Mellon University, Pittsburgh, PA 15213, USA}

\author{\href{https://orcid.org/0000-0001-8868-0810}{Camille Avestruz}}
\affiliation{Department of Physics, University of Michigan, Ann Arbor, MI 48109}
\affiliation{Leinweber Institute for Theoretical Physics, University of Michigan, Ann Arbor, MI 48109, USA}

\author{\href{https://orcid.org/0000-0002-1472-5235}{Jeffrey Regier}}
\affiliation{Department of Statistics, University of Michigan, Ann Arbor, MI 48109}

\author{The LSST Dark Energy Science Collaboration}


\begin{abstract}
Stage-IV dark energy wide-field surveys, such as the Vera C. Rubin Observatory Legacy Survey of Space and Time (LSST), will observe an unprecedented number density of galaxies. As a result, the majority of imaged galaxies will visually overlap, a phenomenon known as blending. Blending is expected to be a leading source of systematic error in astronomical measurements.
To mitigate this systematic, we propose a new probabilistic method for detecting, deblending, and measuring the properties of galaxies, called the Bayesian Light Source Separator (BLISS).
Given an astronomical survey image, BLISS uses convolutional neural networks to produce a probabilistic astronomical catalog by approximating the posterior distribution over the number of light sources, their centroids' locations, and their types (galaxy vs. star).
BLISS additionally includes a denoising autoencoder to reconstruct unblended galaxy profiles.
As a first step towards demonstrating the feasibility of BLISS for cosmological applications, we apply our method to simulated single-band images whose properties are representative of year-10 LSST coadds.
First, we study each BLISS component independently and examine its probabilistic output as a function of SNR and degree of blending.
Then, by propagating the probabilistic detections from BLISS to its deblender, we produce per-object flux posteriors.
Using these posteriors yields a substantial improvement in aperture flux residuals relative to deterministic detections alone, particularly for highly blended and faint objects.
These results highlight the potential of BLISS as a scalable, uncertainty-aware tool for mitigating blending-induced systematics in next-generation cosmological surveys.
\end{abstract}

\section{Introduction}
The distribution of matter in the universe reflects the underlying cosmology that governs its evolution.
We can probe the matter distribution through cosmic shear, in which foreground matter coherently distorts background galaxies due to weak gravitational lensing. Cosmic shear is a powerful probe that can be used to constrain cosmology \citep{huterer2002weak, kilbinger2015cosmology} and has been applied successfully in cosmological surveys, including the Sloan Digital Sky Survey (SDSS; \citealt{huff2014shear}), 
Dark Energy Survey (DES; \citealt{secco2022desy3_shear,amon2022_desy3_shear_calibration, desy6_shear}), 
Subaru Hyper Suprime-Cam (HSC; \citealt{dalal2023hsc_shear_spectrum,li2023hsc_shear_2pcf}), and 
Kilo-Degree Survey (KiDS; \citealt{asgari2021kids}).
Stage-IV dark energy surveys, such as the Vera C. Rubin Observatory Legacy Survey of Space and Time (LSST; \citealt{lsst2009book, lsst2019overview}), will observe an unprecedented number density of sources because of their greater depth than earlier wide-field surveys. 
State-of-the-art surveys such as LSST, through analyses of cosmological probes such as cosmic shear, will enable unprecedented precision and accuracy in estimates of cosmological parameters. However, cosmological constraints remain limited by systematics arising from astrophysical processes and measurement procedures \citep{weinberg2013cosmic, mandelbaum2018weak}.

In cosmological analyses based on galaxy catalogs derived from optical surveys, a leading systematic is \textit{blending} \citep{melchior2021challenge}. 
Blending refers to the visual overlap of light sources in astronomical images, which biases and degrades galaxy detection and measurement of their properties.
For example, blending can impact the distribution of measured galaxy shapes \citep{hoekstra2017study} and photometry \citep{song2018photometry}, the selection function of sources \citep{hartlap2011bias}, and the estimated redshift distribution of galaxies $n(z)$ \citep{maccrann2022desy3}.

Blending already represents a significant source of systematic error in weak lensing measurements in current ground-based Stage-III surveys \citep[\eg,][]{maccrann2022desy3}, and it is expected to become more dominant in Stage-IV surveys, given their greater depth and smaller statistical uncertainties.
Simulations suggest that between $62\%$ of galaxies in the LSST will be blended to some degree \citep{sanchez2021effects} and between $15\%$ and $30\%$ of galaxies will be unrecognized blends, in which two or more light sources are identified as one \citep{dawson2016ambiguous, troxel2023joint}. 
Unrecognized blends can significantly affect weak-lensing and cosmological measurements. 
\cite{euclid2019undetected} finds that introducing a distribution of faint galaxies in a simulation of Euclid VIS images creates a multiplicative shear bias of at least a few times $10^{-3}$. 
\cite{nourbaksh2021effects} uses simulations to forecast the effect of unrecognized blends on the derived structured growth parameter $S_8$ for LSST. They find that blending introduces a bias greater than the $2\sigma$ statistical error in estimates of this parameter. 

Deblending methods, also known as ``deblenders'', aim to mitigate this important systematic.
``Deblender'' can refer to methods that perform one of the following functions on an astronomical image containing blended light sources: 
\begin{enumerate}
    \item \textbf{Detection}: Determine the number of light sources in an astronomical image and their centroids.

    \item \textbf{Deblended Measurement}: For each identified source in a blend, estimate the properties of that source while attempting to remove the contributions from nearby sources. 
    
    \item \textbf{Segmentation}: For each identified source in a blend, estimate the flux contribution to each pixel in the image. From this information, individual galaxy properties, such as total flux or shape, can also be estimated.
\end{enumerate}

One of the most widely used tools for identifying light profiles in an image is SourceExtractor (\sextractor) \citep{bertin1996sextractor}, which uses a model-independent image thresholding technique to detect objects \citep{lutz1980algorithm}. It can also perform basic segmentation based on a multiple isophotal analysis technique \citep[based on][]{beard1990cosmos}. One major limitation of \sextractor is that each pixel can belong to only one object. 
The SDSS Photo pipeline \citep{lupton2005sdss} expands upon SourceExtractor, allows for overlap between objects, and estimates the portion of flux in a pixel belonging to each object. 
Neither SourceExtractor nor the SDSS pipeline can account for multiband information. 

Deblenders developed during the last decade, such as MuSCADeT \citep{joseph2016multi} and its successor \scarlet \citep{melchior2018scarlet}, build joint nonparametric models of multiband image data. Here, the light sources are not specified using a parametric model, such as a S\'ersic fit, but rather by an algorithm that enforces constraints on the light distribution. 
In particular, \scarlet can impose an arbitrary number of constraints on each source (e.g., symmetry and monotonicity with respect to a source's centroid), enabling modeling of aspects such as distinct stellar populations in galaxies. 
More recently, machine-learning-based deblenders have demonstrated high reconstruction accuracy on simulated datasets \citep{arcelin2021deblending, biswas2024madness, sampson2024score, patel2025npe}.

Most deblenders are deterministic and do not account for the inherent uncertainty in the properties of a blended source. 
To overcome this limitation, we introduce a new probabilistic deblender: the Bayesian Light Source Separator (BLISS). BLISS builds on recent ideas in deep generative modeling and variational inference \citep{blei2017variational, kingma2019introduction} to quantify blending uncertainty. Propagating this uncertainty to downstream cosmological analyses has the potential to mitigate blending systematics.

\begin{figure*}[!hbtp]
    \centering
    \includegraphics[width=\textwidth]{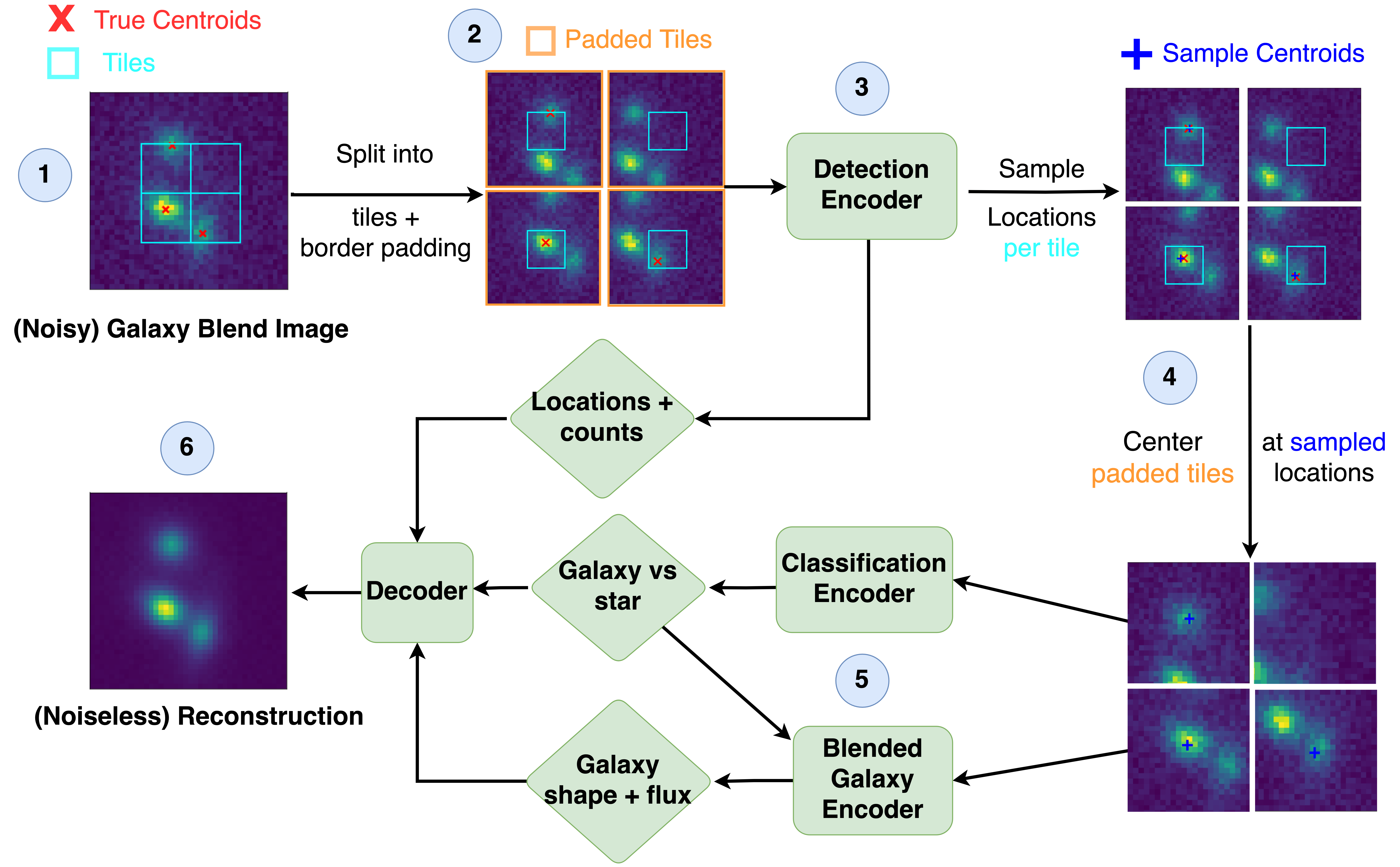}
    \caption{
        \textbf{BLISS inference pipeline outline}. Schematic demonstrating the procedure for detecting, deblending, and reconstructing galaxies with BLISS.
        The procedure begins with an astronomical image of a galaxy blend, which is split into padded tiles (orange border) and passed through the detection encoder. The detection encoder outputs centroid locations for each detected galaxy that can be used to center each padded tile on its corresponding source. The centered padded tiles are then passed to the classification and deblending encoder. Finally, the outputs of each encoder can be used to construct a posterior catalog of the galaxy blend, and a corresponding noiseless reconstruction of the original image. 
        For more details on each of the steps (blue circles) in this outline, see Section~\ref{sec:method}.
    } 
    \label{fig:bliss-outline}
\end{figure*}

BLISS uses simulation-based inference (SBI) to approximate a posterior distribution of deblended galaxy properties (see Section~\ref{sec:method}).\footnote{As explained in Section~\ref{sec:method}, the galaxy deblending component of BLISS is deterministic and uses an autoencoder architecture (Section~\ref{sec:deblender-method}). Despite this, BLISS can produce  posterior distributions of galaxy properties that capture detection uncertainty using the procedure described in Section~\ref{sec:joint-results}.}
BLISS extends the approach from StarNet \citep{liu2023variational}, which uses SBI to detect and measure stars in crowded starfields, to process images containing both stars and galaxies.
Algorithmically, inference involves three encoders and a decoder, as illustrated in Figure~\ref{fig:bliss-outline}. 
After training these models, inference can be performed rapidly on a GPU, in just seconds for megapixel images. 
In this work, we train and evaluate BLISS on simulated single-band images that are representative of typical year-10 LSST wide-field coadd images used for cosmological analyses.

In Section~\ref{sec:dataset}, we describe the image simulations we used to train and evaluate our model. 
In Section~\ref{sec:method}, we describe each component of the BLISS pipeline and explain how it is trained using SBI. 
In Section~\ref{sec:results}, we evaluate the BLISS framework using our simulations. Each component is first evaluated separately; then, in Section~\ref{sec:joint-results}, we present results on combining the probabilistic detection and deblending capabilities of BLISS.
Finally, we conclude in Section~\ref{sec:conclusion}.

\section{Simulated Images} \label{sec:dataset}

We designed our dataset of simulated images to mimic year-10 LSST single-band coadds. The galaxies and stars in our image simulations are modeled using parametric light profiles. The source density is taken from previous LSST simulation work \citep{connolly2014catsim, abolfathi2021dc2}, and we do not model clustering.
All our images are simulated using the \galsim Python package \citep{galsim2015}.\footnote{\url{https://github.com/GalSim-developers/GalSim}} 

\subsection{Survey Settings}
Our studies target a single band, the $i$-band, at the LSST survey pixel scale of 0.2 arcseconds. We use the same parametric point spread function (PSF) model $\pi$ for all our galaxy images, which consists of an atmospheric Kolmogorov component and an optical obscured Airy pattern. We use a FWHM of $0.79$ for the atmospheric component, corresponding to the median zenith seeing for this filter \citep{ivezic2019lsst}.
We add Gaussian noise to our images, assuming the background-dominated limit, so that their SNR resembles that of LSST year-10 images produced by coaddition. 
The added noise corresponds to a sky level $b$, which is chosen to be the average expected sky level for LSST year 10. 
However, the sky level is not added directly to the images; all simulated images are background-subtracted. 
Our setup closely resembles the simulations used in previous blending work focused on LSST \citep{sanchez2021effects}. 
The exact survey parameters are sourced from the LSST module of \surveycodex.\footnote{\surveycodex is a Python library designed to archive the parameters of various surveys with references. Many of its parameters were obtained from the \texttt{WeakLensingDeblending} package \citep{sanchez2021wld}, in some cases with corrections. This library can be found in the following URL: \url{https://github.com/LSSTDESC/surveycodex}}

\subsection{Galaxy Properties}

The galaxy properties for our image simulations come from a galaxy catalog produced with \catsim, an early version of an end-to-end LSST simulation framework \citep{connolly2014catsim}. The galaxies produced by this framework cover the redshift range of $0 < z < 6$. They are simulated using a parametric model for the light profile of galaxies consisting of bulge and disk components.\footnote{The disk component corresponds to a S\'ersic profile with index $n=1$ (also known as an exponential profile), and the bulge component corresponds to a S\'ersic profile with index $n=4$ (also known as a ``de Vaucouleurs'' profile) \citep{Sersic1963influence}.}
The model parameters are the total flux, the relative flux of each component, and the sizes and shapes of each component. The two components share the same centroid and orientation. While \catsim also includes an AGN component, we omit it to simplify the modeling of galaxy morphology. 
The fluxes, shapes, and sizes of galaxies follow a realistic distribution designed to match LSST's key observables, and are informed by previous SDSS observations as well as galaxy formation literature \citep{bruzual2003stellar, de2006formation, gonzalez2009testing}. 

For our study, we used the same one-square-degree subset of the \catsim catalog as in \cite{sanchez2021effects}. This subset of galaxies comprises approximately $850$k galaxies with a limiting $r$-band magnitude of $28$. 
For our datasets, we impose a magnitude cut in the $i$-band of $27$, which corresponds to galaxies with a median SNR of approximately $3$. This yields approximately $576$k galaxies, which we split into three equal sets for training, validation, and testing.

\subsection{Star Properties}

Our datasets include stars to evaluate our deblender in the context of star-galaxy blends. 
The star distribution and fluxes are derived from the LSST DESC Data Challenge Two (DC2) simulation catalogs \citep{abolfathi2021dc2}. 
Specifically, we use the same star catalog as in previous work on cosmic shear measurements in LSST-like simulations \citep{sheldon2023metadetection}. We set our star density to the average over simulated fields, while rejecting fields with density higher than $100\, \text{arcmin}^{-2}$. This results in a star density of $15.5\, \text{arcmin}^{-2}$. 
We apply a lower limit magnitude cut of $20$ and an upper limit magnitude cut of $27$ in the $i$-band, resulting in approximately $17$k stars. The former cut excludes stars with very high SNRs of over $1000$. We model stars as a point source convolved by the same PSF $\pi$ used by galaxies. We do not simulate saturation or other kinds of image artifacts. 

\subsection{Datasets for model training and evaluation}

We create four different types of synthetic datasets to train and test our algorithms. The first dataset, \dtsingle, is used to train a model to reconstruct individual galaxy images (Section~\ref{sec:deblender-method}).
The second dataset, \dtblend, is used to train and evaluate the detection and classification BLISS models (Section~\ref{sec:detection-classification-method}) and to evaluate the deblending model (Section~\ref{sec:deblender-results}). 
The third dataset, \dttiles, is used to train the deblending model (Section~\ref{sec:deblender-method}).
Finally, the fourth dataset is a simplified version of \dtblend designed to evaluate BLISS's joint detection and deblending capabilities (Section~\ref{sec:joint-results}).
These datasets are as follows:

\begin{enumerate}
    \item \dtsingle: Consists of galaxies simulated individually and centered on a square image of size $53 \times 53$ pixels. Each galaxy is convolved with the same parametric PSF $\pi$ described above. We simulate every galaxy in our magnitude-selected sample of the \catsim catalog and split them into three equal parts for training, validation, and testing.

    \item \dtblend: Consists of galaxy and star blends on an image of size $98 \times 98$ pixels. The number of sources in each stamp is Poisson distributed using a mean density of $160 \, \rm{arcmin}^{-2}$ for galaxies and $15.5 \, \rm{arcmin}^{-2}$ for stars.\footnote{These source densities correspond to the galaxies and stars drawn in the image simulation, not the density of detectable sources.}  
    Sources are uniformly distributed throughout the stamp; we do not model clustering.
    We create sets of $10$k, $5$k, and $15$k blends by sampling with replacement from the training, validation, and testing galaxy catalogs in \dtsingle, respectively. Star magnitudes are sampled with replacement from the DC2 star catalog across the three datasets.

    \item \dttiles: Uses the same properties and source density as \dtblend, except that images have dimensions of $53 \times 53$ pixels and the central $5 \times 5$ region always contains exactly one galaxy whose centroid is uniformly distributed within this region. 
    For each image of this dataset, an additional image is created of the same size and noise realization, but where the central galaxy is redrawn to be exactly centered in the image and no other sources exist outside the central $5 \times 5$ region.\footnote{The additional image in this dataset allows us to train the deblender encoder without needing to interpolate images on the fly in a GPU.}
    For this dataset, only training and validation sets of 50k and 10k images, respectively, are created, using the same galaxies as in \dtblend.

    \item \dtcentral: Same as the primary images in the \dttiles, except that images have dimensions $73 \times 73$ pixels, the central galaxy is always exactly in the middle of the image, and all sources are galaxies. Only a test set of $10$k images is created using the corresponding testing galaxies.

\end{enumerate}

\section{Method} \label{sec:method}

BLISS estimates the posterior distribution of light source properties from a given astronomical image, following the procedure laid out in Figure~\ref{fig:bliss-outline}.
We start by decomposing the inference problem into subproblems involving overlapping image subregions, which we call ``padded tiles'' (Section~\ref{sec:tiling}). These padded tiles are passed into a set of three neural network encoders that are trained via variational inference \citep{ambrogioni2019forward} to return the parameters of distributions characterizing the light source properties in each tile (Section~\ref{sec:vi-in-bliss}).
First, the detection encoder returns posterior distributions on the number of objects per tile and the positions of objects' centroids. Sampled centroids from these distributions are used to center the padded tiles (within a pixel, via cropping).
Next, the classification encoder takes these centered padded tiles and returns a probability characterizing the type of each source (star or galaxy) (Section~\ref{sec:detection-classification-method}).
Finally, the deblending encoder uses these centered padded tiles to return (deterministic) latent galaxy representations that can be used to obtain noiseless reconstructions of deblended galaxy images (Section~\ref{sec:deblender-method}). 
Once the three encoders have been trained, BLISS can process astronomical images of any size with the same PSF and resolution as the training images.

\subsection{Tiling \label{sec:tiling}}

The tiling mechanism within BLISS allows for tractable inference of probabilistic catalogs from astronomical images. The tiling procedure used by BLISS encoders exploits the fact that the dependence among well-separated and non-overlapping light sources in a given image can be ignored in the variational distribution when measuring their properties for cataloging.
We split images containing astronomical light sources into overlapping square subregions that we call \textit{padded tiles}; these are input independently to BLISS's encoders. We constrain each encoder to output a prediction exclusively for the source whose centroid lies in a small, central square region of each padded tile.\footnote{By construction, the centroid of a given star or galaxy is contained in a unique tile. Therefore, sources are not double-counted.} We refer to this region as a \textit{tile}.
We illustrate the tiling mechanism with the orange (padded tiles) and blue (tiles) squares in steps 1 and 2 of the schematic of BLISS predictions (Figure~\ref{fig:bliss-outline}). 

BLISS, by construction, outputs a prediction for at most one source within a given tile.  As a result, if the centroids of two or more sources are present in a single tile, BLISS will output an incomplete catalog for that tile. In particular, if at least one source in the tile is detectable, BLISS will output a set of properties corresponding to a single ``hybrid'' source.\footnote{At training time, BLISS always targets the properties of the brightest source present in a tile.} 
We partially mitigate this by using small tiles of size 5 pixels, which reduces the probability that more than one source centroid lies in any given tile. 
Given a source density of $175.5\, \rm{arcmin}^{-2}$, this probability is approximately $0.0012$.
To be concrete, we expect that approximately $10\%$ of images in our \dtblend dataset will have at least one tile with two source centroids. 
In the majority of these occurrences, however, one of the sources involved will be below the detection threshold.\footnote{Realistic levels of clustering might significantly impact these assumptions. Clustering increases the number of close pairs of objects at a given number density \citep{euclid2019undetected, maccrann2022desy3}. This would increase the proportion of objects that fall in the same tile, thus degrading BLISS performance in the manner mentioned above.}

Despite this limitation, the BLISS framework can still detect the vast majority of blended galaxies. As illustrated in Figure~\ref{fig:bliss-outline}, both padded tiles and tiles contain flux from multiple light sources when dividing the image of a blend, even if their centroids land in different tiles. BLISS encoders explicitly account for this in their training procedure. The encoders are trained to remove contributions from light sources in nearby tiles when predicting a source's properties in a given tile. For all encoders, the padded tile size must be large enough to capture the morphological information of galaxies relevant to the corresponding tile. For our datasets, we find $53$ pixels to be sufficiently large.

\subsection{Variational inference\label{sec:vi-in-bliss}}

BLISS adopts a standard Bayesian model of astronomical images, whereby light source properties are latent random variables, and pixel intensities are observed random variables \citep[cf.][]{brewer2013probabilistic, portillo2017improved,regier2019approximate,buchanan2023markov,stone2023astrophot}. Appendix~\ref{app:bayesian-framing} provides details of our specific Bayesian model.

To perform posterior inference under this model, we use variational inference to train the BLISS encoders to approximate the posterior distribution of light source parameters. Specifically, we approximate the true posterior $P(\Zv \vert \Xv)$ of a light source catalog $\Zv$ given an astronomical image $\Xv$ with an approximate posterior distribution, known as the \textit{variational distribution}. We choose our variational distribution to factorize over tiles:
\begin{equation}
    Q_{\vphi}(\Zv) = \prod_{t=1}^{T} Q_{\vphi}(\Zv_{t}),
\label{eq:tile-factorization}
\end{equation}
where $\Zv_{t}$ is the catalog for tile $t$.
We implicitly condition each factor on $\Xv_{t}$, the padded tile corresponding to tile $t$. The sub-index $\vphi$ corresponds to the weights of the neural network that takes as input a padded tile $\Xv_{t}$ and outputs the parameters characterizing the distribution $Q_{\vphi}(\Zv_{t})$ for tile $t$.

The factorization over tiles used in Equation~\ref{eq:tile-factorization} simplifies the requirements of our encoders and allows us to account for the variable number of light sources in a given image. 
Our fitted variable distribution becomes a better approximation of the posterior distribution as the tiles become smaller and padded tiles become larger. 
Smaller tiles reduce the likelihood of multiple source centroids landing in the same tile. 
Larger padded tiles are more likely to contain enough information to characterize the source whose centroid lies in the tile.

In our approximation, the inferred tile catalog $\Zv_{t}$ contains the light source properties $\{n_{t}, \ellv_{t}, b_{t}, \vec{z}_{t}\}$, where $n_{t} \in \{0, 1\}$ is the number of sources whose centroids lie in the tile, $\ellv_{t} \in [0, 1]^{2}$ is the centroid position of the tile's source in units of tile length,  $b_{t} \in \{0, 1\}$ is the classification of the source as a galaxy ($b_{t} = 1 $) or star ($b_{t} = 0$), and $\vec{z}_{t} \in \mathbb{R}^{8}$ is a latent vector characterizing galaxy morphology and brightness (for more details, see Section~\ref{sec:deblender-method}).

We explicitly define the variational distribution $Q_{\vphi}(\mathcal{Z}_{t}) = Q_{\vphi}(n_{t}, \ellv_{t}, b_{t}, \vec{z}_{t})$ for each tile $t$ as follows:
\begin{align}
    n_{t} \sim \text{Bernoulli}(\omega_{n, t})\label{eq:counts-variational} \\ 
    \left[ b_{t} \mid n_{t}=1 \right] \sim \text{Bernoulli}(\omega_{g, t}) \label{eq:bernoulli-variational} \\ 
    \left[ \ellv_{t} \mid n_{t}=1 \right] \sim \mathcal{N}(\vec{\mu}_{\ell, t}, \text{diag}(\vec{\nu}_{\ell, t})) \label{eq:centroid-variational} \\
    \left[\vec{z}_{t} \mid n_{t}=1, b_{t}=1 \right] \sim \delta_{\vec{\zeta}_{t}}
\end{align}
where $\text{diag}(\cdot)$ is used to denote a diagonal matrix, $\mathcal{N}(\mu, \Sigma)$ denotes a normal distribution, and $\delta_{x}$ represents a delta function centered at $x$. The distributional parameters,
\begin{equation}
    \vec{\theta}_{t} = \{\omega_{n,t}, \omega_{g,t}, \vec{\mu}_{\ell,t}, \text{diag}(\vec{\nu}_{\ell,t}), \vec{\zeta}_{t}\},
\end{equation}
are the output of neural networks with weights $\vphi$ applied to a given padded tile $\Xv_{t}$. These weights are the variational parameters. We model locations using a bivariate Gaussian in the variational distribution. Samples from this distribution that lie beyond the bounds of a given tile are considered non-detections.\footnote{When the true centroid is close to the tile boundary or highly uncertain, there is a significant probability that the posterior samples fall outside the tile. We choose to discard these samples instead of assigning this source to a separate tile due to our current simplifying assumption that tiles are completely independent. Future work will explore sharing information between neighboring tiles to improve the variational approximation.
}
Additionally, we chose the variational distribution for the latent galaxy representation $\vec{z}_{t}$ to be a delta function. This implies that we only return point estimates for this quantity with no associated uncertainty.

Critically, our method is \textit{amortized} \citep{kingma2013auto,rezende2014stochastic}: the variational parameters $\vphi$ are shared among tiles (i.e., $\vphi$ is not indexed by $t$); a single set of neural networks with weights $\vphi$ is used for any padded tile of any image. If our method were not amortized, as in more classical variational inference approaches \citep{blei2017variational}, we would need to run an iterative optimization procedure for each tile, which would make our procedure highly inefficient. 
Amortization allows us to train a single set of encoders only once, which greatly speeds up inference for large datasets.

In the next two sections, we describe the numerical optimization procedure that we use to obtain the neural network weights $\vphi$ that determine our variational distribution $Q_{\vphi}$ for any given padded tile $\Xv_{t}$. We describe the procedure to optimize the detection and classification encoders in Section~\ref{sec:detection-classification-method}, and the procedure for training the deblending encoder in Section~\ref{sec:deblender-method}.

\subsection{Detection and classification encoders} \label{sec:detection-classification-method}

BLISS uses forward amortized variational inference (FAVI) \citep{ambrogioni2019forward} to perform posterior inference for light source properties other than galaxy parameters. Two of the BLISS encoders are optimized using FAVI: the detection and classification encoders.
As presented in Equations~\ref{eq:counts-variational} through \ref{eq:centroid-variational}, the detection encoder outputs the parameter of a Bernoulli distribution describing whether a source is present in a given tile. It also outputs two parameters describing where the source centroid is within the tile.
The classification encoder outputs a parameter indicating whether a source is a star or a galaxy. 

We now describe our procedure for training these encoders.  In this section, $\Zv$ refers to the light source parameters captured by the detection and classification encoders, so that:
\begin{equation}
    \Zv \equiv \{n_{t}, b_{t}, \ellv_{t}\}_{t=1}^{T}.
\label{eq:partial-catalog}
\end{equation}

To find the optimal weights $\vphi$ for our neural networks, we minimize 
\begin{equation}
    L(\vphi) = \mathbb{E}_{\Xv \sim P(\Xv)} \KL{P(\Zv \vert \mathcal{X})}{Q_{\vphi}(\mathcal{Z})},
\label{eq:initial-vi-loss}
\end{equation}
where $\mathbb{E}$ denotes the expectation operator\footnote{The expectation operator is defined by $\mathbb{E}_{\xv \sim P(\xv)}f = \int f(\xv) P(\xv)\,d\xv$, for a generic function $f$ and a random variable $\xv$ with probability distribution $P$.} and $\KL{\cdot}{\cdot}$ the KL divergence, which, for generic distributions $p$ and $q$, is
\begin{equation}
    \KL{p}{q} \equiv \int p(z) \log \left( \frac{p(z)}{q(z)} \right) dz.
\label{eq:kl-divergence}
\end{equation}
The KL divergence quantifies the difference between two distributions. In particular, if $p$ and $q$ are equal almost everywhere, then $\KL{p}{q} = 0$. 

Using a specific factorization for the variational distribution and dropping terms that do not depend on $\vphi$, we can rewrite the loss function from Equation~\ref{eq:initial-vi-loss} as 
\begin{align}
    L'(\vphi) = - \mathbb{E}_{\mathcal{X, Z} \sim P(\mathcal{X, Z})} \sum_{t=1}^{T} & \Big[ \log Q_{\vphi}(n_{t}) + \log Q_{\vphi}(\ellv_{t} \vert n_{t}) \nonumber \\ 
    &+ \log Q_{\vphi}(b_{t} \vert n_{t}, \ellv_{t}) \Big].
\label{eq:vi-loss}
\end{align}
For the full derivation, see Appendix~\ref{app:vi-derivation}. With this loss function, we train our detection and classification encoders, which are parameterized together by neural network weights and biases $\vphi$. Each encoder outputs a subset of the distributional parameters $\vec{\theta}_{t}$ for each tile.

Note that the evaluation on padded tiles implicitly treats the problem as independent sub-problems in that the model outputs independent posterior approximations for each tile. However, the inputs to the encoders are padded tiles, which are not independent of each other. The padding in a given padded tile includes pixels also contained in nearby padded tiles, which means that each encoder uses the information from nearby light sources when inferring light source parameters in a given tile.

To form unbiased estimates of the loss function (Equation~\ref{eq:vi-loss}), we perform the following steps:
\begin{enumerate}
    \item[1.] Randomly sample a mini-batch of size $B$ from our training dataset of galaxy and star blends \dtblend (Section~\ref{sec:dataset}), with images $\{\Xv_{b}\}_{b=1}^{B}$ and full catalogs $\{\Zv_{b}\}_{b=1}^{B}$ .

    \item[2.] Divide each image into padded tiles $\{\Xv_{b, t}\}_{t=1}^{T}$, and obtain the corresponding tile catalogs $\{\Zv_{b, t}\}_{t=1}^{T}$ from the full catalog.\footnote{In the case that more than one source centroid is present in a given tile of an image, we only keep the brightest source present in that tile for the corresponding tile catalog. This ensures our training procedure is well-defined but might bias our algorithm, as discussed in Section~\ref{sec:tiling}.} 

    \item[3.] Use a neural network with weights $\vphi$ that takes the padded tiles as input and outputs distributional parameters $\vec{\theta}_{t}$, which define the variational distributions of each tile.

    \item[4.] For each variational distribution on each tile, evaluate its density on the corresponding true source parameters of that tile. The sum of relevant variational distributions corresponds to the loss of each tile.

    \item[5.] Sum the loss over all tiles of each image in each mini-batch.

    \item[6.] Average the loss over the mini-batch examples to evaluate the expectation. 
    Finally, the gradient of this loss is propagated to update $\vphi$.
\end{enumerate}

We train two independent encoder networks corresponding to the detection and classification encoders. 

First, the detection encoder takes padded tiles as inputs and outputs distributional parameters corresponding to source counts and centroids. On each tile $t$, this encoder outputs 1) the probability of a source being present $\omega_{n,t}$ and 2) the mean $\vec{\mu}_{\ell,t}$ and the diagonal of the covariance matrix $\vec{\nu}_{\ell,t}$ for its centroid. 
These distributional parameters define a variational distribution density, which is then evaluated on true source counts and centroids to obtain the loss. 
Training this encoder only uses the first two terms of Equation~\ref{eq:vi-loss}.\footnote{For all encoders, tiles not containing a light source do not contribute to their loss function, except when outputting detection probability via the term $Q_{\vphi}(n_{t})$.}

Next, the classification encoder outputs $\omega_{g,t}$ for each padded tile to parameterize the Bernoulli distribution in Equation~\ref{eq:bernoulli-variational}. Only the last term of the loss function (Equation \ref{eq:vi-loss}), $\log Q_{\vphi}(b_{t} \vert n_{t}, \ellv_{t})$, is used for training this encoder.
The training of the classification encoder is similar to that of the detection encoder, but with one difference: the inputs to the classification encoder are padded tiles that have been centered (within a pixel) on the source in the tile by cropping pixels around its true centroid location. This is illustrated by the image on the right in Figure~\ref{fig:bliss-outline}. 
We found that this centering step increased the performance of the classification encoder. 
Tiles that do not contain sources ($n_{t}=0$) do not contribute to the loss function of the classification encoder. 

Finally, we optimize both neural networks with a stochastic gradient procedure using the Adam optimizer \citep{kingma2014adam}. We use a learning rate of $10^{-4}$ and otherwise default parameters of the \pytorch implementation \citep{pytorch2019}. We use mini-batches of size $32$ from \dtblend.\footnote{We chose this combination of parameters because doing so leads to a smooth and rapid decline in the loss function curve during training. The hyperparameters of other encoders in this work are chosen similarly.} 
The neural network architecture for these encoders follows a standard ResNet-like architecture \citep[e.g.,][]{he2016deep} consisting of convolutions, batch norms, dropouts, and ReLUs.\footnote{The exact implementation of our neural network can also be found in our public repository: \url{https://github.com/prob-ml/bliss/tree/desc-oja}.}

\subsection{Deblending encoder} \label{sec:deblender-method}
We train the last of the BLISS encoders, the deblending encoder, using a two-step approach that is similar to the procedure in \cite{arcelin2021deblending}. 

First, we train an autoencoder (AE) model to capture a nonparametric description of the individual galaxies we simulate with \galsim. Our AE consists of two pairs of neural networks: an encoder and a decoder. The encoder neural network summarizes the noisy image of an isolated and centered galaxy $\xv$ with a latent representation $\vec{z} \in \reals{8}$, which captures both the flux and morphology of the noiseless galaxy image.\footnote{The $8$ numbers that compose the latent vector $\vec{z}$ do not individually have a physical meaning. This vector can be thought of as a ``summary'' produced by the encoder network that contains the information required to reconstruct galaxy light profiles contained in the training dataset.} The decoder takes this latent representation as input and outputs a noiseless reconstruction $\tilde{\xv}$ of the original galaxy image. 

We train our AE using the negative log-likelihood between the noisy image of a centered individual galaxy $\xv$ and its reconstruction $\tilde{\xv}$ under the Gaussian approximation to the Poisson distribution:
\begin{equation}
    L(\xv, \tilde{\xv}) = - \sum_{p \in \, \text{pixels}} \log \mathcal{N}(x_{p}; \tilde{x}_{p}, b),
\label{eq:log-normal-likelihood-loss}
\end{equation}
where $\mathcal{N}(x; \mu, \sigma^{2})$ refers to evaluating the density of a normal distribution $\mathcal{N}(\mu, \sigma^{2})$ at the point $x$. The sky level, or background, $b$ is assumed to be known and constant across all simulated images. The AE is trained on individual galaxy images in batches of size $128$ from the \dtsingle dataset (Section~\ref{sec:dataset}) using the Adam optimizer with a learning rate of $10^{-5}$ and otherwise default parameters.

Next, we train the deblending encoder. The deblending encoder is functionally similar to, and uses the same neural network architecture as, the encoder component of the galaxy model AE. The primary difference between the two is that the deblending encoder is trained on blended galaxies, and thus learns to ignore the flux from galaxies overlapping with the target one.
The deblending encoder targets the same loss function as the AE (Equation~\ref{eq:log-normal-likelihood-loss}), and training is performed using samples from \dttiles (Section~\ref{sec:dataset}).
First, images with a shifted central galaxy are centered (within a pixel) at the true centroid by cropping surrounding pixels, resulting in images with size $49 \times 49$. 
These centered images are passed to the encoder, which produces a latent vector $\vec{z}$ capturing the properties of the central galaxy in this image. 
Then, we apply the decoder model (with frozen weights) from the trained AE to create a reconstruction $\tilde{\vec{x}}$ from each of these encodings. The galaxy centroids in these reconstructions are exactly centered in the images.
Finally, we compute the loss in Equation~\ref{eq:log-normal-likelihood-loss} by using the additional redrawn, centered image included in \dttiles as the target for the reconstruction $\tilde{\vec{x}}$.

We train the deblending encoder using the \adam with a learning rate of $10^{-4}$ on batches of blends from the \dtblend dataset of size $128$. The network architecture is the same as that of the AE, and both consist of a series of convolutions, leaky ReLUs, and fully connected feedforward layers. 
Once trained, our deblending encoder can output a latent representation of each detected galaxy in an image.

Finally, we emphasize the distinct statistical approaches to detection/classification and inferring galaxy representations. In the detection and classification encoders, each light-source parameter is associated with a variational distribution factor that captures the uncertainty in that parameter. In contrast, the deblending encoder returns only point estimates with no uncertainty. 
Despite this, we can construct posterior catalogs on galaxy properties that capture detection uncertainty by using samples from the detection encoder as inputs to the deblending encoder. 
This procedure will be demonstrated in Section~\ref{sec:joint-results}.
Future work could use a VAE \citep[as in][]{arcelin2021deblending} or FAVI \citep{patel2025npe} to capture the full uncertainty in the properties of blended galaxies.

\section{Results} \label{sec:results}

We evaluate the performance of BLISS on galaxy blend images rendered with \galsim using the set of $15$k blends from the test part of \dtblend described in Section~\ref{sec:dataset}. 
First, we define the signal-to-noise ratio (SNR) and blendedness of galaxies in Section~\ref{sec:snr-bld}.
Next, we evaluate each of the three encoders separately in Sections~\ref{sec:detection-results} through \ref{sec:deblender-results}. 
Then, in Section~\ref{sec:toy-blend-results} we explore the uncertainty predictions of our encoders in a controlled setting where we vary the distance between a pair of galaxies.
Finally, in Section~\ref{sec:joint-results} we explore how the probabilistic output of BLISS can be used to improve aperture flux recovery.

\subsection{SNR and Blendedness} \label{sec:snr-bld}

In our experiments, we use the signal-to-noise ratio (SNR) and blendedness ($B$) to characterize sources when evaluating the performance of BLISS. We define these quantities below. 

We use the aperture photometry module in SEP, a Python implementation of SourceExtractor \citep{bertin1996sextractor, barbary2016sep}, to measure the signal-to-noise ratio (SNR) defined as
\begin{equation}
    \sqrt{\sum_{p \in \text{aperture}} \frac{I_{p}^{2}}{\sigma^{2}_{n, p}}},
\end{equation}
where $I_{p}$ is the intensity of the image in pixel $p$, and $\sigma_{n, p}^2$ is the sky noise variance. 
We assign an SNR to individual galaxies. When galaxies are blended, we draw each member individually and use this equation to compute its \textit{true SNR} (see Section~\ref{sec:detection-results} for details). 
We use an aperture radius of $5$ pixels throughout our experiments for all photometry measurements.

We adopt the definition of blendedness from \cite{bosch2018hyper}. For a galaxy with intensity $\iv$, \textit{blendedness} is
\begin{equation}
    B = 1 - \frac{\langle \iv, \iv \rangle}{\langle \iv, \iv_{\rm total} \rangle},
\end{equation}
where $\iv_{\rm total}$ is the total intensity from every source in the blend and, for intensities $\vec{I}_{1}$ and $\vec{I}_{2}$,
\begin{equation}
    \langle \vec{I}_{1}, \vec{I}_{2} \rangle = \sum_{p \in \text{pixels}} I_{1, p} I_{2, p},
\end{equation}
where the sum is over all pixels in the image.
In all our experiments, blendedness is always computed using the intensities obtained from the true noiseless images of individual galaxies.
By construction, $B \in [0, 1]$, with $B=0$ corresponding to an isolated galaxy and $B=1$ to a completely overlapping and subdominant galaxy.
Galaxies with higher blendedness will tend to be more difficult to detect and measure accurately compared to galaxies with low blendedness.

\subsection{Detection evaluation} \label{sec:detection-results}

We evaluate the trained detection encoder on galaxy blends and compare it with SEP.
Specifically, we look at the precision, recall, and $F_{1}$ score of the detection encoder using different probability thresholds for what constitutes a detection. 
For the BLISS detection encoder, for each tile having a detection probability larger than a given threshold, we choose the centroid prediction to be the mean of the predicted centroid posterior on that tile.
For SEP, we use the configuration described in Appendix~\ref{app:sep-settings}, which is designed to match a previous blending study \citep{sanchez2021effects}. We do not attempt to vary the SEP parameters.

We first define \textit{matched detections} as detections that are within $2$ pixels of a true centroid \textbf{and} assigned to a unique true centroid using the Hungarian matching algorithm \citep{kuhn1955hungarian} (available as the Scipy function \texttt{linear\_sum\_assignment}). This is a bipartite graph matching algorithm that minimizes a certain cost function on the graph edges. 
The cost function is
\begin{equation}
    \sum_{i} \sum_{j} C_{i,j} X_{i, j},
\end{equation}
where the sum is over all possible matches (between true source $i$ and detection $j$) and $X_{i,j}$ is $1$ if $i$ is matched to $j$ and $0$ otherwise. Let $r_{i,j}$ be the Euclidean pixel distance between the centroid of source $i$ and the detection $j$, then we define $C_{i,j}$ as
\begin{equation}
    C_{i, j} = \begin{cases}
        r_{i,j} \text{\;if\;} r_{i,j} < 2 \text{\, pixels} \\ 
        \infty \text{\; else \;}
    \end{cases}
\end{equation}
Next, we define the \textit{true SNR} of a galaxy as the SNR computed using the true centroid on an image redrawn with the same pixel-noise as the original blended image, but where only this galaxy is present. Similarly, we define the \textit{detected SNR} for a given detection as the SNR computed at this location in the given blended image via aperture photometry. 

Finally, for a given SNR bin, precision is defined as the number of matched detections whose detected SNR lies within this bin divided by the number of detections whose detected SNR is in the bin. Similarly, recall is the number of matched detections for which the matched galaxy's true SNR lies within this bin, divided by the number of galaxies whose true SNR lies in the bin. The $F_{1}$ score is defined as
\begin{equation}
    F_{1} = 2 ( \text{precision}^{-1} + \text{recall}^{-1} )^{-1}
\label{eq:f1-score}
\end{equation}
on a per-bin basis. 

\begin{figure*}[!hbtp]
    \centering
    \includegraphics[width=0.9\textwidth]{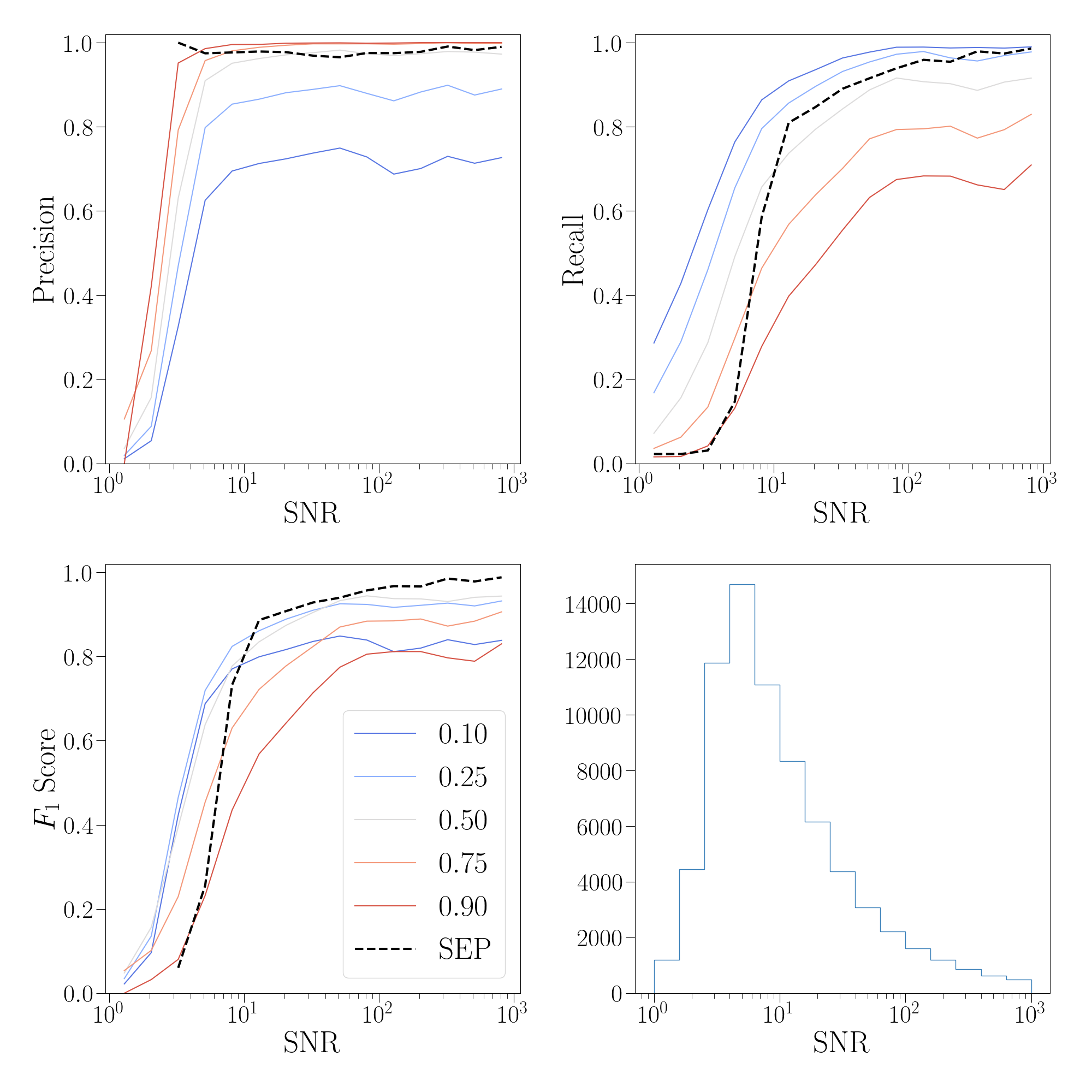}
    \caption{
        \textbf{Detection performance as a function of SNR.} In this figure we present the performance of the trained BLISS detection encoder using different probability thresholds (color gradient, blue to red) and SEP (dashed, black) on our testing set of galaxy blends. We show the precision, recall, and $F_{1}$ score in the same equally-log-spaced SNR bins. In the bottom-right plot, we additionally show the true SNR distribution of our \dtblend dataset using the same SNR bins. For more details on this figure, see Section~\ref{sec:detection-results}.
    } 
    \label{fig:detection-snr-results}
\end{figure*}

Figure~\ref{fig:detection-snr-results} shows the precision, recall, and $F_{1}$ score on shared SNR bins of the detection encoder using different probability thresholds (color gradients, blue to red) and SEP (dashed black). We also show the SNR distribution in the bottom-right panel. The SNR bins in all plots are equally spaced from $3$ to $1000$ on a log scale. 
In the precision plot, we see that the redder curves have higher overall precision than the bluer curves for the detection encoder. This is expected, as a higher probability threshold corresponds to a stricter requirement on the confidence of the BLISS model on detections. 
In comparison, SEP appears to have higher precision than most BLISS probability thresholds across all SNR bins. 
In the recall plot, we observe the opposite trend. A lower probability threshold for the BLISS detection model corresponds to higher recall, as more true objects are found. SEP has relatively higher recall than the detection encoder, except for the low ($<10$) SNR bins. 
The $F_{1}$ score shows that SEP has the most comparable performance to the detection encoder at the fiducial probability threshold ($0.5$).
Finally, as the SNR histogram shows, the bins with the highest performance (approximately above SNR of $50$) do not correspond to the vast majority of galaxies that will be used for cosmological analyses (around $10$ SNR).

\begin{figure}[!hbtp]
    \centering
    \includegraphics[width=0.45\textwidth]{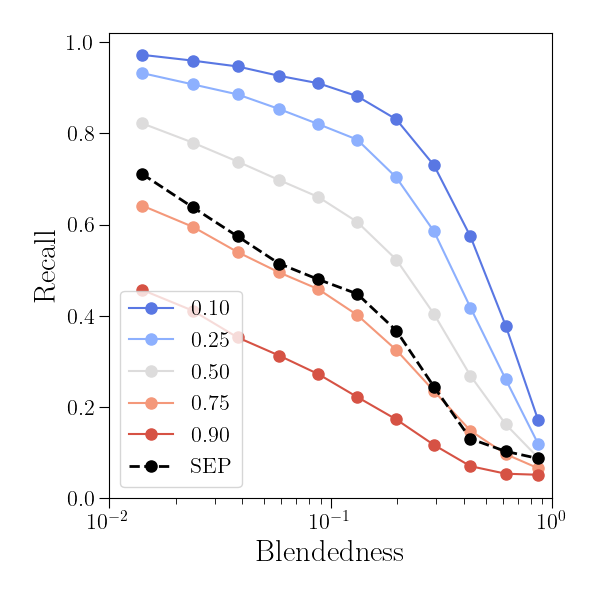}
    \caption{
        \textbf{Completeness as a function of blendedness.} This figure shows the fraction of detected true objects (i.e., recall) as a function of their blendedness for SEP and the BLISS detection encoder.  
        The BLISS detection encoder's recall for different probability thresholds is shown in the blue to red gradient curves, while SEP's is shown with a black dashed curve. The blendedness bins shown are chosen so that there are the same number of sources in each of them. See Section~\ref{sec:detection-results} for details.
    } 
    \label{fig:detection-bld-results}
\end{figure}

Next, in Figure~\ref{fig:detection-bld-results}, we show the recall of the detection encoder and SEP as a function of blendedness.\footnote{We do not include the other two metrics for blendedness as there is not an obvious definition for ``detected blendedness''.} The blendedness is computed using the true noiseless image of each galaxy, and the recall is defined analogously for blendedness bins as for true SNR bins. In this case, we use blendedness bins containing the same number of sources.
We see that the recall of the detection encoder increases as the probability threshold decreases, as in the previous figure. In this case, the recall of the nominal model (threshold of $0.5$) outperforms SEP in finding galaxies across all blendedness bins.
Additionally, we see that the overall recall is lower in Figure~\ref{fig:detection-bld-results} than in Figure~\ref{fig:detection-snr-results}. 
This is related to the fact that most blendedness bins tend to be dominated by the large number of low SNR sources.

Overall, these results show that, compared to a sensible SExtractor configuration, the BLISS detection encoder can recover a larger number of noisy and blended sources in our image simulations while achieving comparable precision over a wide SNR range.

\subsection{Classification evaluation} \label{sec:binary-results}

In this section, we evaluate the next encoder in the BLISS pipeline, the binary encoder, independently of the detection encoder. We provide the true counts and centroids of light sources to the binary encoder when processing images. 

\begin{figure*}[!hbtp]
    \centering
    \includegraphics[width=0.95 \textwidth]{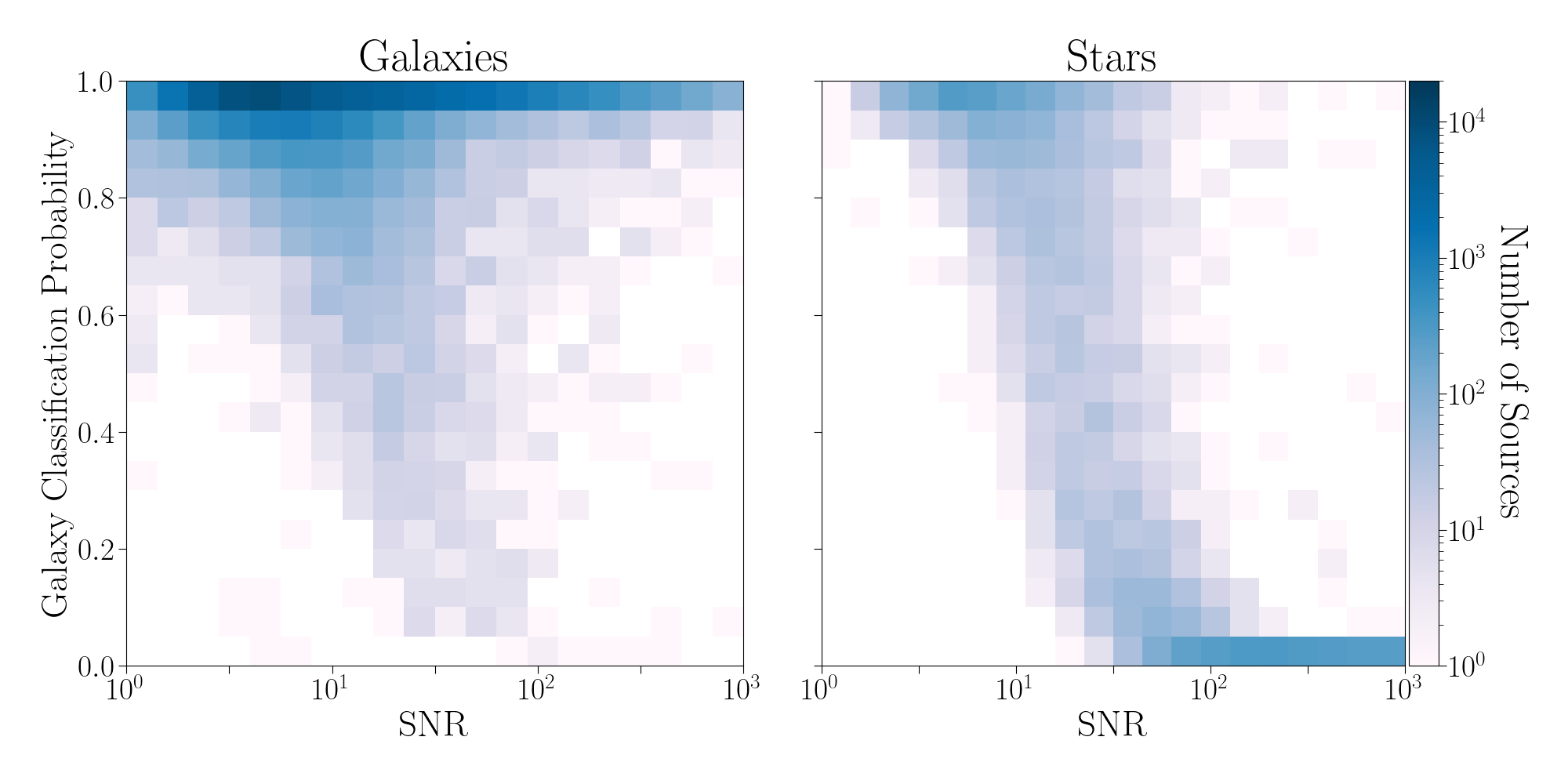}
    \caption{
        \textbf{2D Histogram of classification probability and SNR.} In these plots we show a 2D histogram of the classification probability of a given source being a galaxy and its true SNR, for every galaxy (left) and star (right) in the testing \dtblend. 
        This probability is the output from the binary encoder conditioned on true counts and centroids of every source.
        For more details on this figure, see Section~\ref{sec:binary-results}.
    } 
    \label{fig:binary-contours}
\end{figure*}

In Figure~\ref{fig:binary-contours}, we show 2D histograms of the probability of a given light source being a galaxy outputted by the binary encoder and its true SNR (as defined in Section~\ref{sec:detection-results}). The left histogram corresponds to galaxies, and the right one corresponds to stars.
We see that for high SNRs (above $100$), the model is very confident about the source's type, as expected. As the SNR decreases, the model becomes more uncertain about the classification of any given light source, and we see a vertical strip in both histograms.
As the SNR decreases even further, at about SNR=$10$, we see that the model classifies a large portion of stars as galaxies with high confidence. This occurs because the model defaults to the prior in the case of low SNR sources. Given that the vast majority of sources used during training were galaxies, the model tends to classify very noisy stars as galaxies.

\begin{figure}[!hbtp]
    \centering
    \includegraphics[width=0.45\textwidth]{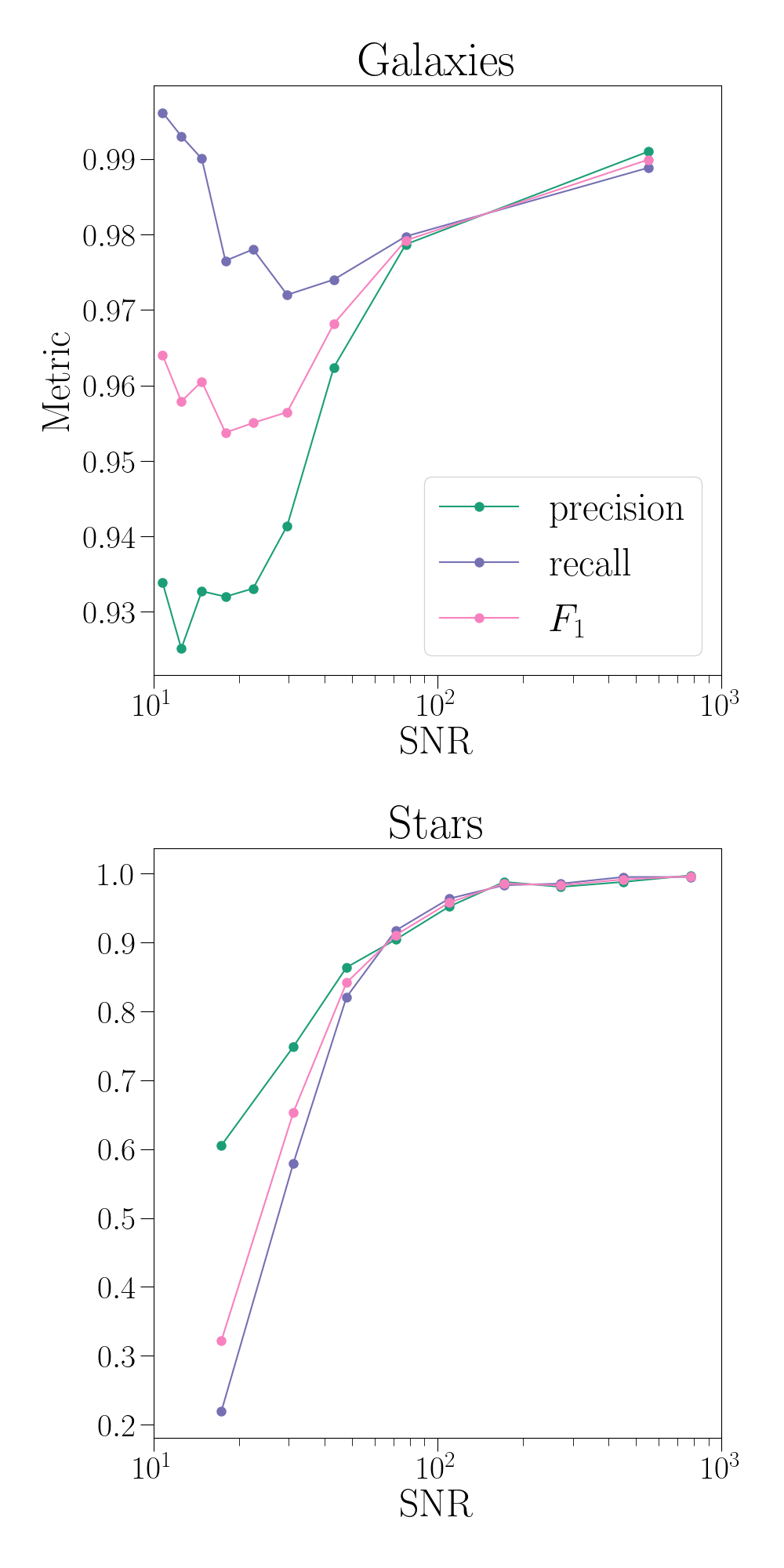}
    \caption{
        \textbf{Classification performance as a function of SNR for galaxies and stars.} This figure shows the precision (green), recall (purple), and $F_{1}$ score (pink) as a function of true SNR obtained by applying the binary encoder to our testing \dtblend of galaxies and star blends. The SNR bins in each case are chosen so that there are an equal number of light sources on each bin. We condition the binary encoder on the true counts and centroids of every source in this dataset. For more details on this figure, see Section~\ref{sec:binary-results}.
    } 
    \label{fig:binary-curves}
\end{figure}

We additionally use the precision, recall, and $F_{1}$-Score metrics, as in the previous section, to evaluate the encoder. These metrics are computed differently for this classification context. 
First, we split these metrics into a galaxy version and a star version, as shown in Figure~\ref{fig:binary-curves}. This is helpful as \dtblend is highly imbalanced, with a much larger number of galaxies than stars. We say that the model classifies a light source as a galaxy if the output probability of the source being a galaxy is $> 0.5$, and a star otherwise.\footnote{The optimal choice for this probability cut will depend on the specific application. The performance demonstrated in Figure~\ref{fig:binary-curves} can be tuned by modifying this value as needed.} 
The precision for galaxies (stars) in this context corresponds to the number of correctly classified galaxies (stars) divided by the total number of sources the model classified as galaxies. The recall for galaxies (stars) is then the number of correctly classified galaxies (stars) divided by the total number of galaxies. The $F_{1}$ score is computed as in Equation~\ref{eq:f1-score}. These quantities are binned and plotted in Figure~\ref{fig:binary-curves} based on the true SNR of the sources, and bins are constructed so that they contain the same number of galaxies or stars. We exclude the faintest sources with $\text{SNR} < 10$ from this figure for visualization purposes. 
In the left plot, we can see that all of the galaxy metrics are above $90\%$ for all relevant SNRs. This is expected given our highly unbalanced dataset. The precision decreases rapidly from around $98\%$ to $93\%$ at the SNR roughly corresponding to the vertical strip in Figure~\ref{fig:binary-contours}. The recall increases as the SNR decreases as the model defaults to the prior and classifies most objects as galaxies. 
Meanwhile, we see that all the star metrics are monotonic as a function of SNR. This is consistent with Figure~\ref{fig:binary-contours} as the purple dots consistently move from the top to the bottom as the SNR increases. 

Our model becomes more uncertain when classifying low SNR sources, as expected. 
However, for very low SNR, the prior dominates, and a significant number of stars are classified as galaxies with high confidence. 
One way to make progress could be to incorporate additional prior information into the network that takes into account the model of stars (i.e., the PSF). For example, using the \textit{spread model} estimator \citep[e.g.,][]{mohr2012spread}.

\subsection{Deblending evaluation} \label{sec:deblender-results}
In this section, we evaluate the deblending encoder independently from the rest of the encoders. We again use \dtblend  and investigate the accuracy of the corresponding galaxy reconstructions. 
We evaluate fluxes measured with aperture photometry, using an aperture of $5$ pixels. The size and ellipticities come from the adaptive moments routine in \galsim.\footnote{We use the \texttt{galsim.hsm} module: \url{https://galsim-developers.github.io/GalSim/_build/html/hsm.html}} 
The size is defined as $T \equiv |\det \bm{M}|^{1/4}$, where $\bm{M}$ is the second-moments matrix. The ellipticity uses the distortion (\texttt{e}) definition from \galsim.\footnote{The definition can be found in: \url{https://galsim-developers.github.io/GalSim/_build/html/shear.html}}
Measurements are made on \textit{residual images} of each galaxy obtained as follows:
\begin{enumerate}
    \item For each galaxy in each blend of \dtblend, we use the deblending encoder to encode this galaxy conditioned on its true centroid. The decoder part of the autoencoder then uses this encoding to create a reconstruction of this galaxy (Section~\ref{sec:deblender-method}). 

    \item Each individually reconstructed galaxy from the decoder is centered on a stamp the size of a padded tile. We use \galsim to interpolate\footnote{We use the default settings for interpolating on \galsim, which uses a ``quintic'' $2$D interpolation for images. See this page for details: \url{https://galsim-developers.github.io/GalSim/_build/html/interpolant.html}} each galaxy to its true location within the padded tile. These padded tiles can be used to create a full reconstruction of the galaxy blend in the original image. 

    \item For each galaxy in a given blend, we use these interpolated reconstructed images to subtract the contribution from all galaxies in the blend except this one. We also remove the contributions from galaxies in the padding and from all stars in the image using their true models. These are the residual images for every galaxy in the dataset. 

    \item Finally, we perform aperture photometry flux measurements and moments, as described above, using the true centroid of each galaxy as input.
\end{enumerate}

We use these residual images for measurements rather than the reconstructed images directly because they contain unphysical artifacts in the tails of galaxies, which can have a large effect on measured moments.
See Section~\ref{sec:toy-blend-results} for additional discussion.

\begin{figure*}[hbt]
    \centering
    \includegraphics[width=0.725 \textwidth]{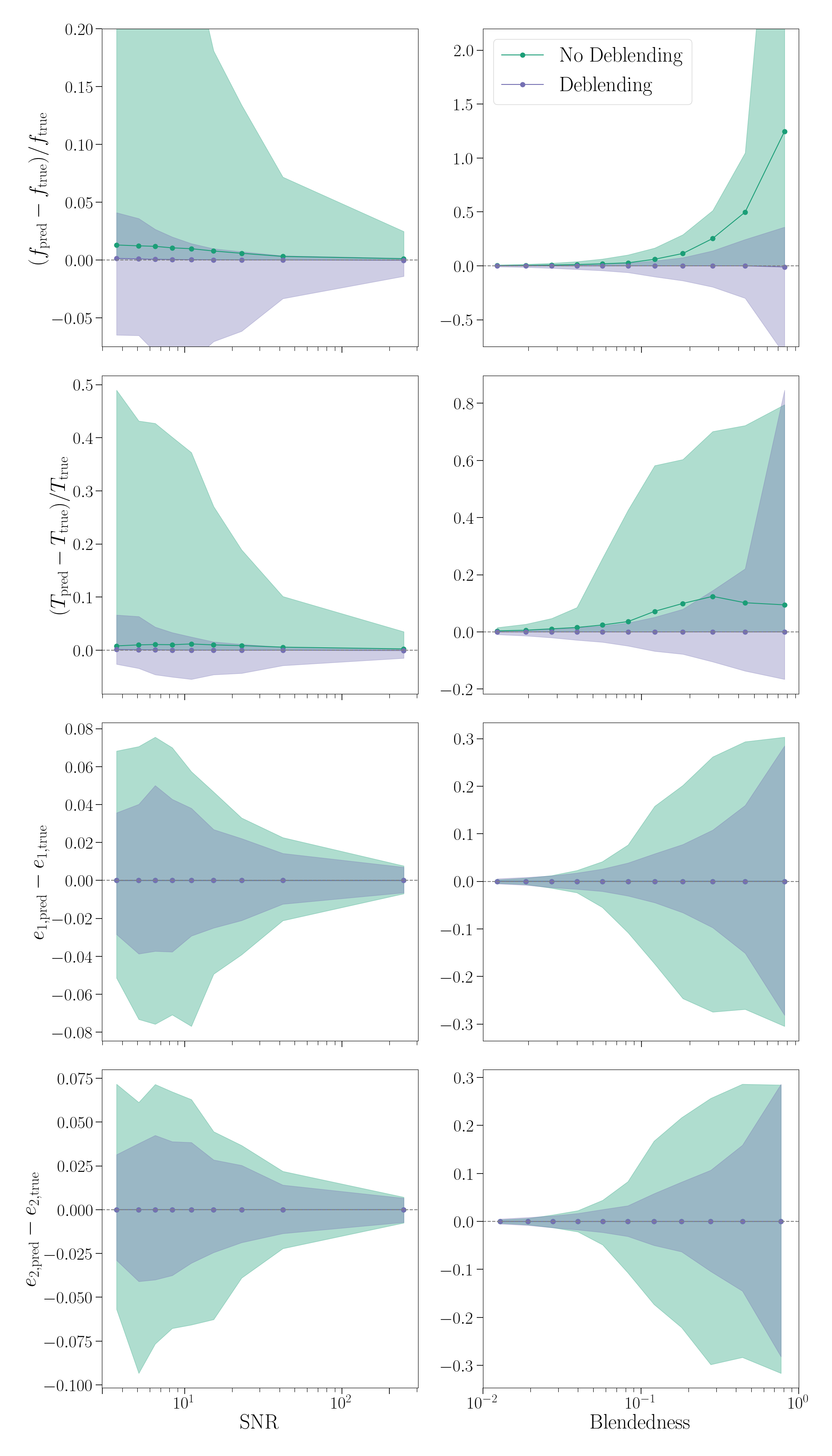}
    \caption{
        \textbf{Residual galaxy measurements as a function of SNR and blendedness.} In this plot we show the median residual flux, size, and ellipticity from measurements on galaxy blended images using true galaxy centroids. The flux ($f$) measurement is performed via aperture photometry with an aperture of size $5$ pixels. 
        The size ($T = |\det \bm{M}|^{1/4}$) and ellipticities ($e_{1}$, $e_{2}$) are measured using the \galsim adaptive moments routine.
        The green color curves corresponds to the measurements with no deblending performed, but using the true centroid of each galaxy. 
        The purple color curves use the reconstructed models from the deblending encoder to remove every other galaxy in the image before performing the measurement. The shaded regions correspond to quantiles of the $1\sigma$ deviation from a Gaussian distribution (i.e., $0.159$ and $0.841$ respectively). For more details on this figure, see Section~\ref{sec:deblender-results}. 
    } 
    \label{fig:deblending-bins}
\end{figure*}

We plot the residuals for fluxes, sizes, and ellipticities obtained using this procedure on every galaxy in our dataset as a function of true SNR and blendedness in Figure~\ref{fig:deblending-bins} (purple curves). 
The green curves (``No deblending'') show these measurements on the same images using true centroids, but no galaxy reconstructions are subtracted before the measurement. All galaxies and stars in the image padding are still subtracted. Finally, the true values of every galaxy measurement are computed by subtracting every other source in the image using the true individual images.
We see from this figure that the scatter in measurements of all quantities is significantly reduced from the deblending. 
In the case of the flux and size measurements, we see that deblending unbiases the residual across all SNR and blendedness bins. 
In the case of no deblending, we see that the flux and sizes medians are always positive, and there is no negative scatter. This is expected, as the flux and size measurements of galaxies can only get larger when flux is added to the image. 

This figure shows that our deblender is capable of reconstructing galaxies accurately, but the biases and errors shown here will be too optimistic as it ignores detection effects.
We investigate the impact of detection on flux and moments measurements in the upcoming sections.

\subsection{Example probabilistic output as a function of distance} \label{sec:toy-blend-results}

In this section, we present examples of the probabilistic output of BLISS on a pair of example galaxies at varying separations.
We note that the exact output of the BLISS encoders on any given galaxy blend can be highly dependent on the images' noise realization. 
Thus, the goal of the results and figures presented in this section is to illustrate aspects of the models' output rather than to draw general conclusions on their performance.

\begin{figure*}[hbt]
    \centering
    \includegraphics[width=0.95 \textwidth]{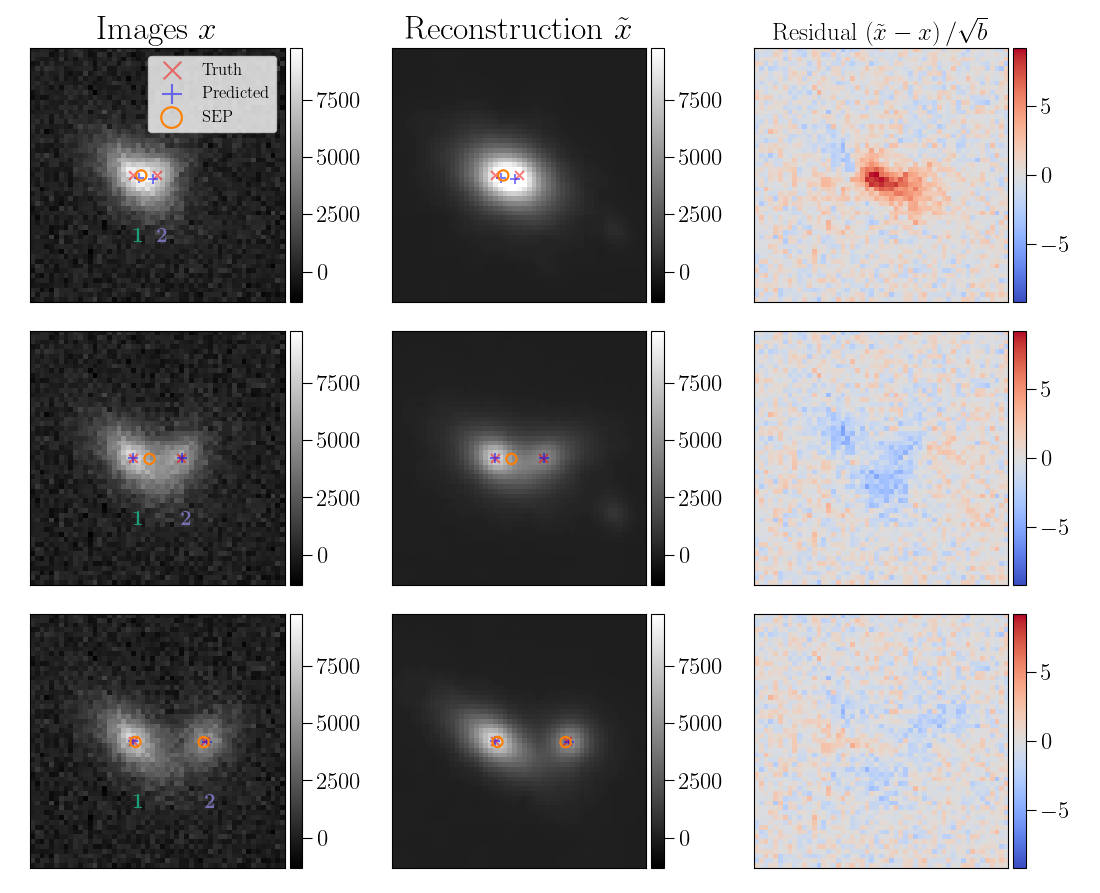}
    \caption{
        \textbf{Image residuals and detections on galaxy pairs at three separations.} In this plot we show the pair of example galaxies used throughout Section~\ref{sec:toy-blend-results} at three different separations: $5$, $10$, and $15$ pixels. The galaxy on the left (labeled with ``$1$'', in green) is fixed at the center of the image. The galaxy on the right (labeled with ``$2$'', in purple) is shifted horizontally at different separations from Galaxy 1. 
        In the first column, we see the true noisy image at each separation with the true centroid of each galaxy in red and the predicted centroid from BLISS in blue. The SEP prediction for centroids is also shown with an orange circle as a reference. 
        The second column shows the noiseless reconstruction from BLISS with the same centroids plotted as in the first column. 
        The last column shows the residual image in units of the noise standard deviation $\sqrt{b}$.
    } 
    \label{fig:toy-image-residuals}
\end{figure*}

Our setup consists of two elliptical exponential galaxies with SNRs of approximately $120$ and $70$ with a variable separation between them as shown in Figure~\ref{fig:toy-image-residuals}.\footnote{These relatively high SNR galaxies are chosen to isolate the impact of blending from that of missed detections of low SNR sources.}
The left galaxy, denoted ``Galaxy 1'' and labeled with a green $1$ in this figure, is fixed at the center of the image. The right galaxy, denoted ``Galaxy 2'' and labeled with a purple $2$ in this figure, shifts horizontally between $0$ and $20$ pixels from the center of the image in intervals of $0.1$ pixels. These results are based on $200$ images, each $103 \times 103$ pixels, with the same Gaussian noise realization added to all images.
The specific properties of these galaxies can be found in Table~\ref{tab:toy-blend-parameters}. 

\renewcommand{\arraystretch}{1.3}
\begin{table}
\centering
\begin{tabular}{||c | c | c||}
\hline
Property & Galaxy 1 & Galaxy 2  \\ [0.5ex]
\hline \hline
$f$ & $\power{2}{5}$  & $\power{1}{5}$ \\
\hline
SNR & 120 & 70 \\
\hline
$a_{d}$ & 1.5 & 1.0 \\
\hline
$q_{d}$ & 0.7 & 0.7   \\
\hline
$\beta$ & $\pi/4$ & $6\pi/4$ \\
\hline
\end{tabular}
\caption{
    The true galaxy parameters of the galaxies in the pair blend from Figure~\ref{fig:toy-image-residuals}. Both galaxy light profiles are exponential. The symbols are as follows: $f$ is the flux, $a_{d}$ is the semi-major axis of the disk, $q_{d}$ is the ratio of the semi-major and semi-minor axes, and $\beta$ is the inclination angle.
}
\label{tab:toy-blend-parameters}
\end{table}

In Figure~\ref{fig:toy-image-residuals}, we show the image of the galaxy pair at three different separations ($5$, $10$, and $15$ pixels) in the first column. The second column contains the noiseless reconstruction from the BLISS deblending encoder (using the predicted centroid as input), and the third column shows the residual image. We also overplot the true centroids and the predictions from SEP and the BLISS detection encoder (using the centroid posterior mean). The figure shows that the detection encoder correctly identifies the total number of galaxies, even at small pixel separations when they visually look like a single blob. As the separation between the objects increases, we see that detection accuracy also increases, as expected. The SEP settings used in this figure are the same as those in Section~\ref{sec:detection-results} and are listed in Table~\ref{tab:sep-settings}.

Next, in terms of reconstructions and residuals, the third column shows image residuals (in units of the noise standard deviation) improving as the separations increase. At the lowest separations, BLISS misestimates the flux from both galaxies as well as the total flux of the system. This may be because the deblending encoder's reconstructions for each tile are computed independently, without any constraint to preserve the total flux of the blend; when galaxies are severely blended, the deblender might include or remove too much flux from each independent prediction.
Finally, we note that all noiseless reconstructions in the second column have noticeable non-physical features, especially in the tails of galaxies. The deblending encoder does not impose any particular galaxy model or physical constraints on its reconstructions, which could explain these unphysical features. 
This effect does not impact our conclusions significantly since we perform measurements only on images with noise added (see Section~\ref{sec:deblender-results}), which buries these features.
Both of these observations suggest avenues for future work in the deblender (Section~\ref{sec:conclusion}).

\begin{figure}[!h]
    \centering
    \includegraphics[width=0.45 \textwidth]{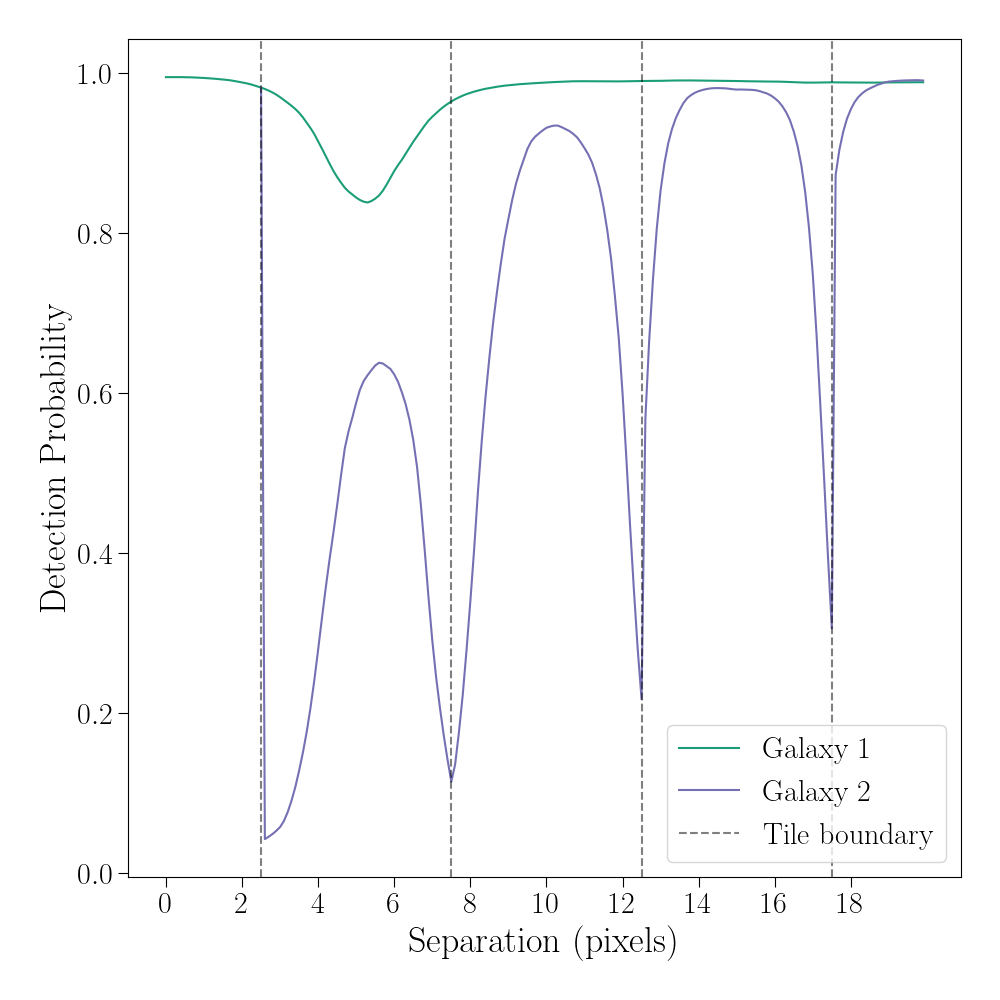}
    \caption{
        \textbf{Detection probability of each galaxy as a function of separation.} This plot shows the detection probability that BLISS assigns to the tile containing Galaxy 1 (green) and 2 (purple) as a function of the separation between them (as described in Figure~\ref{fig:toy-image-residuals}). 
        The dashed vertical lines correspond to the separations at which Galaxy 2's centroid lands in a tile boundary. See Section~\ref{sec:toy-blend-results} for more details on this figure.
    } 
    \label{fig:toy-detection-prob}
\end{figure}

Next in Figure~\ref{fig:toy-detection-prob}, we show the detection probability as a function of the separation between two galaxies. The detection probability for each galaxy is obtained by taking the output of the detection encoder for the tile containing the true centroid of that galaxy. We see that the detection probability for Galaxy $1$ remains relatively stable and does not drop below approximately $0.85$. However, the detection probability of Galaxy $2$ varies in more interesting ways. There are peaks and valleys roughly corresponding with the center and boundaries (dashed lines) of tiles, respectively. 
The second galaxy's detection probability can drop below $0.5$ depending on how close its centroid is to the tile boundary, which is an undesirable artifact of our tiling approach. However, the overall probability curve increases as the two sources become more separated, as expected.
This effect seems to artificially reduce the detection probability regardless of the brightness of the source, and could cause detectable sources to go undetected by our algorithm.  
We discuss possible mitigation strategies in Section~\ref{sec:conclusion}.

In Appendix~\ref{app:pair-blend-centroids}, we also explore the quality of centroid predictions from the detection encoder on these images.
In the next section, we jointly evaluate the detection and deblending encoders in the context of flux measurements on a large sample of blended galaxies.

\subsection{Joint Detection and Deblending Uncertainty} \label{sec:joint-results}

One advantage of BLISS over traditional approaches is the ability to combine probabilistic detection with source deblending. BLISS could be particularly powerful in the context of unrecognized blends, where the total number of sources is ambiguous. 
In this section, we quantify the improvements in photometric measurements achieved by this approach using a large dataset of simulated blended galaxies.

We created a new dataset, \dtcentral (Section~\ref{sec:dataset}), comprising $10$k images for this experiment. 
This dataset is equivalent to the one used in Sections~\ref{sec:detection-results} through \ref{sec:deblender-results}, except that no stars are included in the simulation, and, in every image, the central $5\times5$ tile always contains exactly a single galaxy with an $i$-band magnitude brighter than $25.3$. The image size in this dataset is slightly smaller ($73 \times 73$ pixels) than in \dtblend.
We only evaluate the flux recovery of this central galaxy.
We excluded stars from this simulation to simplify the setup, but the binary encoder could have also been sampled to identify stars. 
We also force the central galaxy to be exactly in the middle of the central tile, which mitigates the effect of multiple sources being present in the same tile (Section~\ref{sec:tiling}), and reduces the boundary effect discussed in Section~\ref{sec:toy-blend-results}.
These simulation choices were made to explore potential improvements to flux measurement via the BLISS probabilistic output while mitigating issues in our current model that will be addressed in future work.
Note, however, that we do not completely eliminate these artifacts as other galaxies in the image could land in the same tile or have their centroid close to a tile boundary. 
Finally, we set a magnitude limit on the central galaxy to ensure its detectability and reduce computational time. 

We compare aperture photometry measurements of the central galaxy in each image using the detection and deblending output from BLISS in three different ways.
The first way, denoted ``MAP'', uses the most likely BLISS detections to reconstruct the light profile of sources and for flux measurement. These detections correspond to those with a tile detection probability larger than $0.5$, and whose location is the mean of the predicted centroid posterior distribution.
The second approach (``SEP'') uses SEP (using the same settings as in Section~\ref{sec:detection-results}) for detections instead of BLISS, but deblending is still performed using the BLISS deblending encoder.
For the third way, denoted ``Samples'', we draw $100$ samples from the BLISS predicted distributions for counts and centroids to obtain a flux distribution by repeating the deblending procedure $100$ times per image. This flux posterior distribution approximately captures the corresponding detection uncertainty in flux measurements. 
We use the mean of this flux posterior distribution as our new point estimator for comparison with the MAP and SEP flux predictions.
For each image, we use the same deblending procedure for aperture flux measurement as in Section~\ref{sec:deblender-results}, where detected galaxies besides the central one are removed using noiseless reconstructions from the deblending encoder. 
The key difference with the previous set of deblending results is that we use the predicted galaxy centroids rather than true centroids for both removing sources and to center the aperture for flux measurement.
In all cases, flux is only measured when a predicted detection is less than $2$ pixels away from the central galaxy. We exclude from the results below the small number (approximately $0.2\%$) of images where at least one of the approaches does not detect and match the central galaxy.\footnote{In the case of the ``Samples'' approach, this means that none of the $100$ detection samples matched the central galaxy.}

\begin{figure}[!hbtp]
    \centering
    \includegraphics[width=0.45 \textwidth]{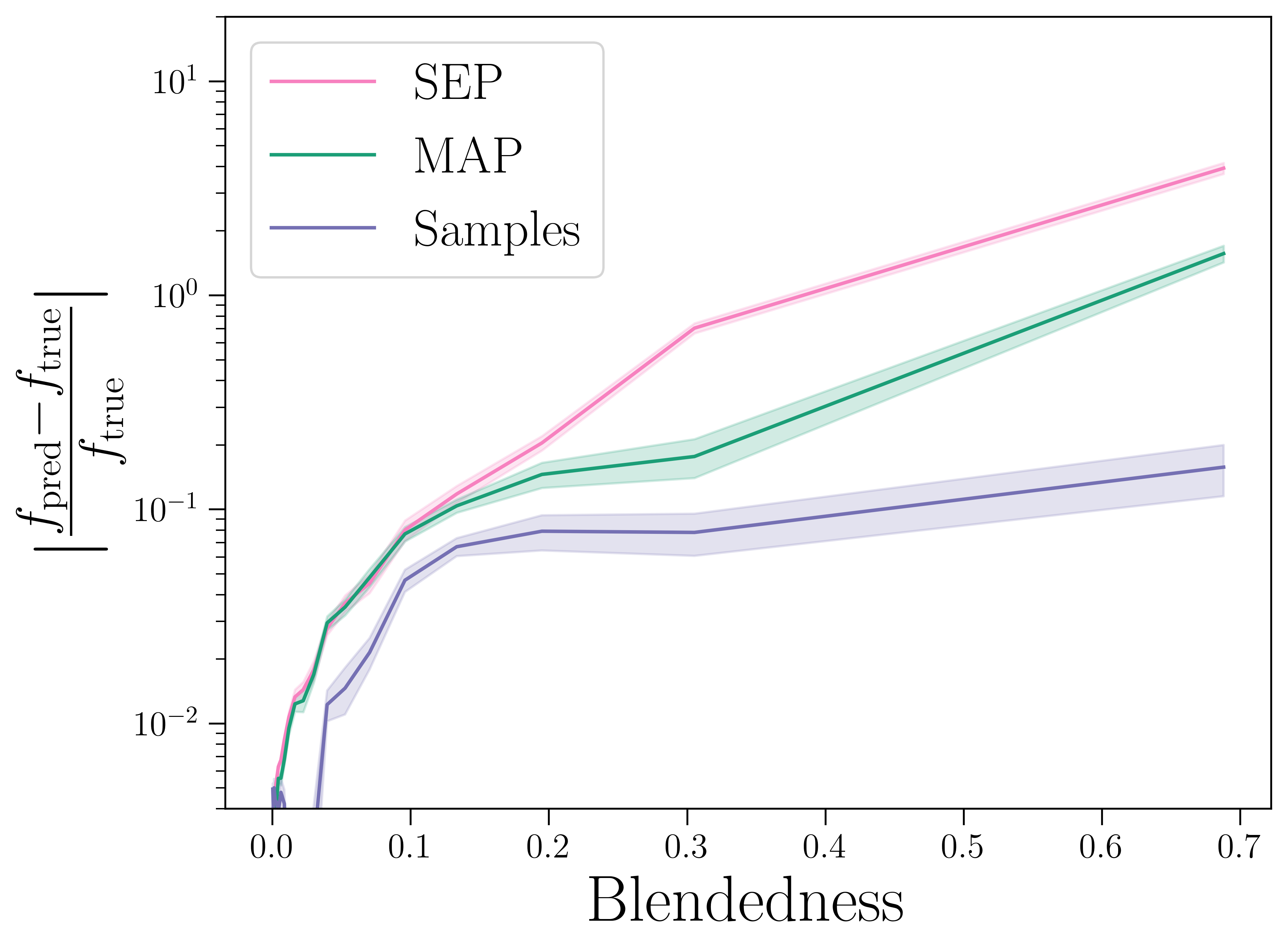}
    \caption{
        \textbf{Flux measurement in BLISS using probabilistic detections as a function of blendedness.} 
        In this figure, we compare the (fractional) absolute median flux residual for three different approaches to flux measurement using BLISS in the dataset \dtcentral. In all three cases, the flux is measured on the central galaxy of each image using aperture photometry after deblending is performed using the same BLISS deblending encoder, but different detections are used. 
        Detections are matched within $2$ pixels of the true galaxy centroid at the center of the image. If no matches with the central galaxy, no flux measurement is performed for that set of detections.
        In the first case (MAP), the most likely detections (above $50\%$ probability) from the BLISS detection encoder are used for deblending and aperture photometry. 
        In the second case (SEP), detections from SEP are used instead. 
        Finally, in ``Samples'', we draw $100$ detection samples from the BLISS detection encoder and use these as input to the deblending encoder. 
        We take the mean over these measured fluxes as the flux estimator for each image. See Section~\ref{sec:joint-results} for more discussion of this figure.
    } 
    \label{fig:samples-blendedness-fluxes-residual}
\end{figure}

Our results are summarized as a function of the blendedness of the central galaxy in Figure~\ref{fig:samples-blendedness-fluxes-residual}. 
We choose a total of $20$ blendedness bins, where $18$ of them are in the range $[3.80\times10^{-4}, 3.76 \times 10^{-1}]$ and are set to ensure an equal number of objects in each. 
Objects with values lower than $3.80\times10^{-4}$ or higher than $3.76 \times 10^{-1}$ are placed in the first and last bin, respectively. The upper bound was chosen to ensure enough statistics on high blendedness examples. Each bin contains at least $400$ corresponding images used to obtain residuals. 
The curves represent the (fractional) median residual in each blendedness bin from each method, where the ``Samples'' approach uses the average of the flux measured across the samples to compute its residual. 
The shaded regions correspond to the $1\sigma$ bootstrap errors on the median obtained by resampling the residuals on each bin.
In this figure, we see that the sampling approach to measure photometry significantly outperforms the point estimations from both ``MAP'' and ``SEP'' approaches across blendedness bins. 
The gap in the median residuals increases with blendedness and is particularly large for the last bin, where the residual jumps from approximately $16\%$ to about $160\%$ for the MAP and about $400\%$ for SEP.

\begin{figure}[!hbtp]
    \centering
    \includegraphics[width=0.45 \textwidth]{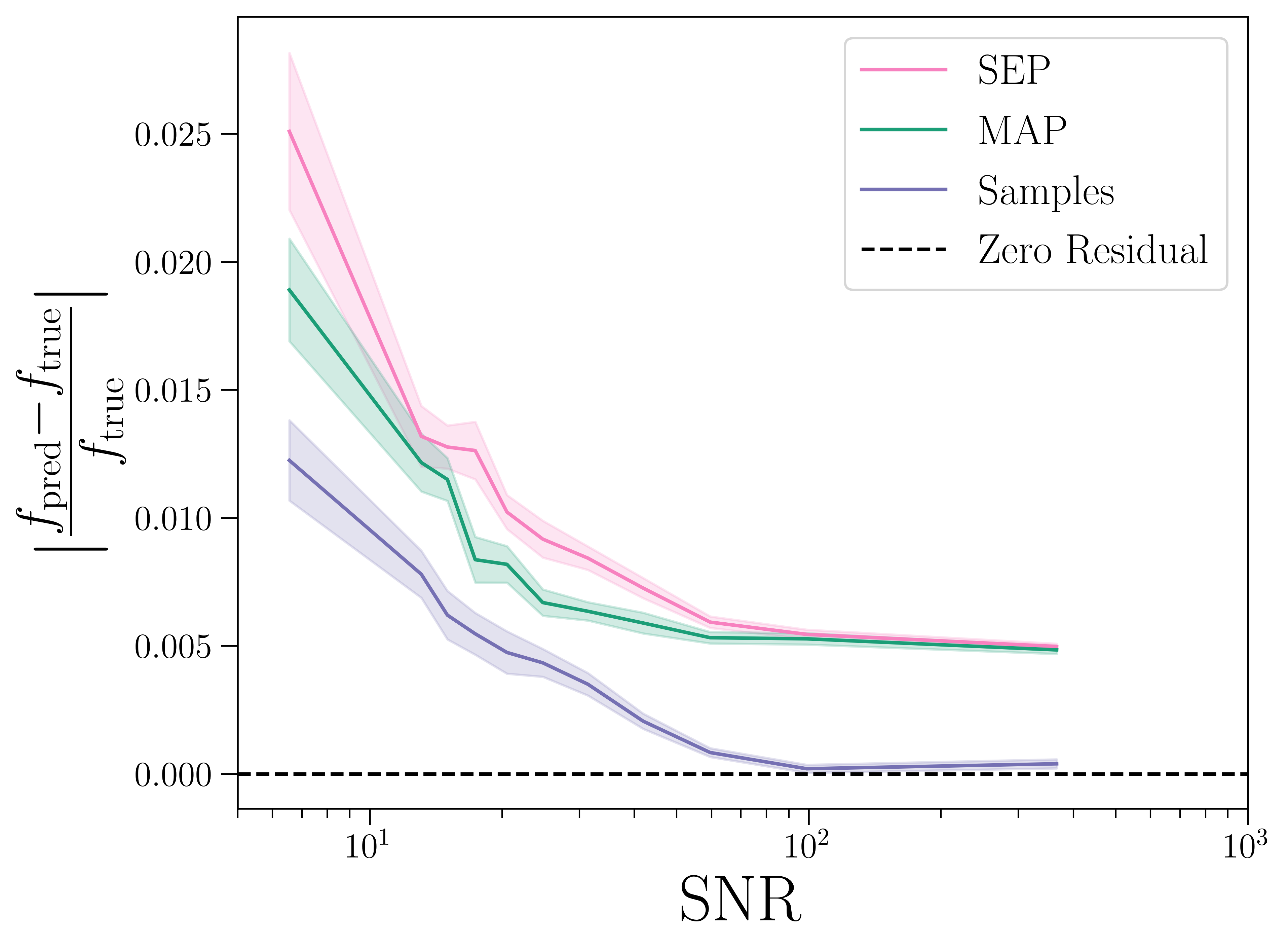}
    \caption{
        \textbf{Flux measurement in BLISS using probabilistic detections as a function of SNR.} 
        This figure presents the same results as Figure~\ref{fig:samples-blendedness-fluxes-residual}, but residuals are split into SNR bins based on the true SNR of the central source of each image. The bins are chosen so that each of them contains the same number of galaxies. The median blendedness value in each bin is $<0.08$, which explains the significantly smaller residual values compared to Figure~\ref{fig:samples-blendedness-fluxes-residual}. See Section~\ref{sec:joint-results} for additional discussion.
    } 
    \label{fig:samples-snr-fluxes-residual}
\end{figure}

In Figure~\ref{fig:samples-snr-fluxes-residual}, we show the flux residual results as a function of the true SNR of the central source of each image. As before, the curves represent the median fractional flux residual and the shading their bootstrap error for the same three methods. 
The SNR bins are chosen to ensure an equal number of images in each, resulting in more than 600 images per bin. 
This figure shows that the sampling approach outperforms the other two methods across the entire SNR range. The residual of all methods increases as the SNR decreases, and the two deterministic methods (SEP and MAP) perform similarly to each other throughout. 
The difference in the magnitude of median residuals between Figure~\ref{fig:samples-blendedness-fluxes-residual} and \ref{fig:samples-snr-fluxes-residual} comes from the fact that the median blendedness across all SNR bins is below $0.085$.

We hypothesize that these improvements are driven by the large number of objects undetected by the ``MAP'' or ``SEP'' approach when the blendedness level is high or the SNR is low. 
Undetected objects will not be removed from the image through deblending, and would thus lead to flux overestimation. 
On the other hand, the sampling approach has the potential to detect more sources as no hard cutoff is used for detection probability. 
Samples corresponding to sources in the image that are difficult to detect could, on average, reduce the flux residual.
It is also possible that the sampling approach is less sensitive to the tile boundary artifact, as the probability of a source close to a boundary appears to be shared between the corresponding tiles \citep{liu2023variational}. 
However, the comparison using SEP detections indicates that this cannot be the sole source of improvement, as the SEP algorithm does not employ a tiling procedure.

Overall, this result demonstrates the potential of our method to mitigate photometric biases in blended systems.

\section{Discussion and Summary} \label{sec:conclusion}
In this work, we introduce the Bayesian Light Source Separator (BLISS), a novel simulation-based inference method for detection, deblending, and measurement of astronomical light sources. 
We have presented the components of BLISS and illustrated its performance on LSST-like simulated images of galaxy blends. 
Our key result is that combining probabilistic detection with AE deblending yields substantial improvements in aperture flux measurements for high-blendness systems (Section~\ref{sec:joint-results}).
Our other results are as follows: 
\begin{itemize}

    \item To enable inference of light source parameters for arbitrarily large images, we split our images into overlapping padded tiles. Our tiling scheme forces the neural network to learn only the details local to each light source. For more details, see Section~\ref{sec:tiling}.

    \item We thoroughly evaluate the point estimates from each of our neural networks. 
    For our particular SEP settings (Appendix~\ref{app:sep-settings}), we find that the detection encoder recovers a larger fraction of very faint and high blendedness sources (Section~\ref{sec:detection-results}). 
    Our classification encoder achieves an $F_{1}$ score in excess of $70\%$ when classifying stars with SNR larger than $12$, and in excess of $95\%$ for galaxies with SNR larger than $10$ (Section~\ref{sec:binary-results}). 
    Finally, our deblender significantly improves the recovered flux and morphology of blended galaxies when the true centroid is known (Section~\ref{sec:deblender-results}).

    \item We carefully examine the probabilistic output of our method in the context of a simple simulation of a pair of galaxies. BLISS is capable of detecting sources even at very low source separations. Detection probabilities overall increase as sources become more separated, as expected. However, detection probability can be degraded when sources get too close to tile boundaries. 
    Finally, we find that individual galaxy reconstructions improve as the separation increases.  See Section~\ref{sec:toy-blend-results}.

    \item The inference step of the BLISS pipeline is fast. Once trained using a single GPU, BLISS can perform inference on megapixel images containing galaxy and star blends in seconds.
\end{itemize}

We use an autoencoder (AE) trained on parametric simulated galaxies as our galaxy model. 
This model enables the training of the deblending encoder and the reconstruction of galaxies (Section~\ref{sec:deblender-method}). 
This approach follows a growing body of work on using deep neural networks for galaxy image modeling. Recent work on using deep learning for galaxy modeling commonly uses generative adversarial networks (GANs) and variational autoencoders (VAEs), and, more recently, score-based models \citep{ravanbakhsh2016enabling,dia2019galaxy,fussell2019forging,reiman2019deblending,lanusse2020deep,arcelin2021deblending, hemmati2022deblending, smith2022realistic, biswas2024madness, sampson2024score}.
We used an AE in BLISS since autoencoders are generally more stable in training than GANs \citep{kingma2019introduction} and have been shown to perform galaxy image reconstruction accurately \citep{lanusse2020deep, arcelin2021deblending}.
\cite{lanusse2020deep} used VAEs to build deep generative models that can recreate realistic galaxy images using data from the HST COSMOS catalog \citep{mandelbaum2018HSC}. 
VAEs have also been used successfully for deblending \citep{arcelin2021deblending}. 
These types of methods have the potential to mitigate the model bias often incurred by more standard parametric galaxy models, as demonstrated by \cite{remy2022model}. 
Currently, we do not fully leverage the modeling capabilities of deep generative models in BLISS, as our AE is deterministic and is trained exclusively on parametric galaxies. However, we plan to incorporate realistic galaxy profiles and use a VAE to capture the full uncertainty of galaxy properties in future work. 

BLISS includes a detection encoder (Section~\ref{sec:detection-classification-method}) that is capable of outputting a posterior distribution for the centroids of stars and galaxies in astronomical images.
Traditional detection approaches are based on peak finding or thresholding, such as the popular algorithm \sextractor \citep{bertin1996sextractor, barbary2016sep}. 
However, classical approaches to cataloging cannot characterize uncertainty in the number of sources found and are known to result in a significant number of ambiguously blended sources \citep{dawson2016ambiguous}. 
Probabilistic approaches for detection relying on MCMC have been proposed for deblending crowded starfields, such as in \cite{feder2020multiband}, and predicting point estimates and uncertainties of star fluxes and centroids. 
However, these approaches are computationally expensive because they rely on MCMC.
In this work, we evaluated BLISS point (deterministic) predictions of source detections and found comparable performance with the chosen SourceExtractor configuration (Section~\ref{sec:detection-results}).
However, in this study, we discarded the probabilistic information from the BLISS detection encoder by using thresholds. 
In a later section, we show an example of how this additional information can be used (Section~\ref{sec:joint-results}).

BLISS also includes a classification encoder (Section~\ref{sec:detection-classification-method}) that outputs a probability that a detected source is either a galaxy or a star. 
Approaches to star-galaxy separation have thus far often involved neural networks, such as the one used in \sextractor \citep{bertin1996sextractor}. This neural network differs from ours in that it does not use the images directly but rather 10 carefully chosen numerical features and a feedforward architecture. 
More recent approaches to star-galaxy classification include convolutional neural networks such as BLISS and avoid the need to extract features carefully \citep{kim2016star, garg2022star}. Other approaches have used low-dimensional embeddings of optical images and Gaussian processes for classification \citep{goumiri2020star, muyskens2022star}. 
The CNN architecture used in BLISS could enable the use of information from multiple bands or exposures for a given light source. These approaches will become more important as new surveys observe fainter objects, where the increase in blending of smaller galaxies makes classification difficult \citep{slater2020morphological}.

BLISS performs deblending using its deterministic deblending encoder (Section~\ref{sec:deblender-method}), which takes as input samples from the detection and classification encoders. 
The deblending encoder consists of a neural network that learns to target the galaxy at the center of an image, while removing the flux of surrounding galaxies. 
This approach to deblending is similar to DebVAdEr \citep{arcelin2021deblending} and MADNESS \citep{biswas2024madness}. 
Other deblenders, such as \scarlet \citep{melchior2018scarlet}, use constrained optimization to deblend light sources without deep learning.
In future work, we also plan to conduct an extensive comparison of BLISS with these and other deblenders using the BlendingToolKit package \citep{mendoza2025blending}.

As explored in Section~\ref{sec:toy-blend-results}, our tiling procedure (Section~\ref{sec:tiling}) has potential drawbacks. 
We found that detection probability correlates strongly with distance from the source to the tile boundary. 
This makes the detection probability drop below $0.5$ even for detectable and isolated sources.
This effect could be impacting the results of Section~\ref{sec:detection-results}, degrading detections and reducing the BLISS recall across all SNR and blendedness bins.
In a precursor paper to BLISS, which uses a similar detection encoder architecture \citep{liu2023variational}, the authors find that the probability is ``shared'' between tiles when the source becomes too close to the tile boundary. 
This suggests that drawing samples from the detection encoder might be more robust to the tile boundary effect, as all tiles with significant assigned probabilities can contribute. 
Other ways to mitigate this artifact include allowing neighboring tiles to overlap, or modifying the variational distribution (Equation~\ref{eq:tile-factorization}) to encode dependencies between neighboring tiles \citep{regier2025autoregressive}.
Future work will involve a more in-depth analysis of tile boundary effects and exploring these and other mitigation strategies.

In Section~\ref{sec:joint-results}, we quantified the effect on photometric measurements of combining the probabilistic detection and deblending capabilities of BLISS. 
We generate aperture flux posterior distributions of individual deblended galaxies by repeatedly sampling galaxy centroids from the detection encoder and using these samples as input to the deterministic deblender encoder. 
These flux posterior distributions primarily capture the contribution from the detection uncertainty of the target and nearby sources. 
This procedure could be applied to produce posteriors for any measurable galaxy property, although we only explored improvements to aperture fluxes in this work.
Future work will replace the deblender encoder with a VAE to better approximate the full uncertainty present in blended galaxy systems.
We find that using these deblended flux posteriors can drastically reduce the median flux residual for highly blended galaxies in our slightly simplified \dtcentral simulations (Figure~\ref{fig:samples-blendedness-fluxes-residual}).
While out of scope for this paper due to our use of single-band image simulations, future work using multi-band images could investigate downstream applications to improve color and redshift estimation by deriving their corresponding posteriors.
Previous work has already shown that a machine learning deblender can be used to improve photometric redshift estimation in simulations \citep{merz2025deepdisc}.

Future work on BLISS will focus on preparing this algorithm for use with real astronomical data and supporting cosmological analyses from stage-IV optical surveys such as LSST. Specifically, we plan to: 
(1) Extend BLISS so that processing multi-band data becomes possible. This might enable propagating our predicted flux posteriors (Section~\ref{sec:joint-results}) to photo-z estimation. 
(2) Accommodate data with spatially variable PSFs \citep{patel2025npe} and background, as well as other effects present in real survey images.
(3) Relatedly, training and testing our models using more realistic LSST-like image simulations such as the LSST DESC DC2 simulations \citep{abolfathi2021dc2, duan2025bliss_dc2} and the OpenUniverse 2024 simulations \citep{openuniverse2024}. 
(4) Training and testing our encoders using images of galaxies with realistic morphology. For instance, we could consider combining deep fields and space observations from other cosmological surveys to build a more realistic model of galaxy morphology. In this work, we have used only parametric galaxies, which may introduce model biases in real survey data.
(5) Test and validate our method using a specific set of science-driven metrics. For example, use BLISS detection samples to explore whether shear detection biases could be mitigated.
(6) Address potential biases related to the tiling procedure and corresponding independence assumptions in BLISS (Section~\ref{sec:toy-blend-results}). This includes the fact that we allow at most one source centroid per tile.
(7) Understand how robust BLISS is to significant and relevant differences between the training and test data.
(8) Finally, develop a general methodology that can be used to propagate BLISS predicted uncertainties to downstream cosmological analyses.

\section*{Author Contributions}
IM wrote the manuscript text, developed and maintained the BLISS software version used for this paper, created and co-designed all results figures and the outline figure, implemented star/galaxy classification and galaxy deblending functionality in the codebase, and created datasets for training and evaluation using \galsim. 
DH developed and helped maintain the codebase throughout the analysis and provided feedback on the results. 
RL refactored parts of the codebase, provided guidance on extending StarNet, and offered feedback on the results. 
ZP refactored parts of the codebase to ease development and fine-tuned the detection network to improve performance.
ZZ developed the initial unit tests for the codebase and helped develop a framework for training machine-learning models using PyTorch Lightning.
AG provided guidance on the simulations used, co-designed results figures, and reviewed the manuscript.
CA edited the paper, provided significant comments and suggestions during writing, brainstormed paper organization, and suggested ways to present results. 
JR conceived of the project, guided code development, refactored the codebase, edited the manuscript, and provided feedback on it.
All coauthors regularly discussed BLISS software development and performance evaluation.

\section*{Acknowledgments}

This paper has undergone internal review in the LSST Dark Energy Science Collaboration. We are very grateful to the internal reviewers: Rachel Mandelbaum and Peter Melchior. 
Additionally, we thank Francois Lanusse, Cyrille Doux, and James Buchanan for providing feedback about our results.

This material is based on work supported by the National Science Foundation under Grant No. 2209720 and the U.S. Department of Energy, Office of Science, Office of High Energy Physics under Award Numbers DE-SC0023714 and DE-SC009193.
IM was also supported by the Special Interest Group on High Performance Computing (SIGHPC) Computational and Data Science Fellowship and the Michigan Institute for Computational Discovery and Engineering (MICDE) Graduate Fellowship.
IM and CA also received support from the Large Synoptic Survey Telescope Corporation (LSSTC) Enabling Science Award.

The DESC acknowledges ongoing support from the Institut National de Physique Nucl\'eaire et de Physique des Particules in France; the Science \& Technology Facilities Council in the United Kingdom; and the Department of Energy and the LSST Discovery Alliance
in the United States.  DESC uses resources of the IN2P3 Computing Center (CC-IN2P3--Lyon/Villeurbanne - France) funded by the Centre National de la Recherche Scientifique; the National Energy Research Scientific Computing Center, a DOE Office of Science User Facility supported by the Office of Science of the U.S.\ Department of Energy under Contract No.\ DE-AC02-05CH11231; STFC DiRAC HPC Facilities, funded by UK BEIS National E-infrastructure capital grants; and the UK particle physics grid, supported by the GridPP Collaboration.  This work was performed in part under DOE Contract DE-AC02-76SF00515.

AG acknowledge the support of a grant from the Simons Foundation (Simons Investigator in Astrophysics, Award ID 620789).

We are grateful to the developers of the Python modules that we used:
Astropy \citep{astropy:2013,astropy:2018,astropy2022},
Blending ToolKit \citep{mendoza2025blending},
einops \citep{rogozhnikov2022einops},
GalSim \citep{galsim2015},
Matplotlib \citep{matplotlib2007},
NumPy \citep{numpy2020},
Pytest \citep{pytest},
PyTorch \citep{pytorch2019},
PyTorch Lightning \citep{pytorch_lightning2019},
Ruff\footnote{\url{https://github.com/astral-sh/ruff}},
scikit-learn \citep{scikit-learn},
SciPy \citep{scipy2020},
SEP \citep{bertin1996sextractor, barbary2016sep}\footnote{\url{https://github.com/sep-developers/sep}},
SurveyCodex\footnote{\url{https://github.com/LSSTDESC/surveycodex}},
Tensorboard \citep{tensorflow2015-whitepaper},
Typer\footnote{\url{https://typer.tiangolo.com/}}, and
tqdm \citep{typer}.

\section*{Data Availability}

The code to reproduce all experiments is in the public GitHub repository: \url{https://github.com/prob-ml/bliss/tree/desc-oja}. The software is also available in Zenodo at \url{https://zenodo.org/records/18164685}.


\appendix

\section{Appendix A: Bayesian framework for probabilistic cataloging} \label{app:bayesian-framing}

We describe the Bayesian framework that underpins the problem of probabilistic inference of catalogs from astronomical images. 
Given an image containing blended galaxies and stars, $\Xv$, we want to obtain a probability distribution over catalog $\Zv$. A catalog corresponds to the set of all galaxy and star parameters that characterize the mean intensity of a given image.
Our goal is then to obtain the posterior distribution, $P(\Zv \vert \Xv)$, defined as,
\begin{equation}
    P(\Zv \vert \Xv) = \frac{P(\Zv) P (\Xv \vert \Zv)}{P (\Xv)}.
\label{eq:full-posterior}
\end{equation}
%
Here, $P(\Zv)$ is the prior distribution, $P(\Xv \vert \Zv)$ is the conditional likelihood, and $P(\Xv)$ is the evidence or marginal likelihood. 

The prior $P(\Zv)$ corresponds to an expected distribution of light source parameters before taking any particular astronomical image into account. This distribution could come from astronomical catalogs of precursor surveys or from cosmological knowledge. In our case, the prior over full images $P(\Zv)$ is defined implicitly by our choice of galaxy and star catalogs (see Section~\ref{sec:dataset}). 

The conditional likelihood $P(\Xv \vert \Zv)$ corresponds to the probability that we have observed a specific astronomical image, $\Xv$, given a corresponding catalog $\Zv$. The catalog $\Zv$ describes the source centroid locations and the mean intensity of stars and galaxies (from the flux and shape entries in the catalog). 
The combination of the catalog, the PSF, and the background determines the mean intensity of every pixel in the astronomical image. 
Using the normal approximation to the independent Poisson noise at every pixel of the image, the conditional likelihood reduces to a normal distribution that factorizes over each pixel, where the mean and variance are the mean intensity $\tilde{\Xv}$ determined by the catalog: 
\begin{equation}
    P (\Xv \vert \Zv) = \prod_{p \in \rm pixels} \mathcal{N}(\Xv_{p}; \tilde{\Xv}_{p}, \tilde{\Xv}_{p}).
\label{eq:full-likelihood}
\end{equation}
Though the conditional likelihood is known and tractable to evaluate in our context, we do not use it in our methodology for the detection and classification encoder. We use only the negative log conditional likelihood as the loss function for the galaxy autoencoder and deblending encoder (see Section~\ref{sec:deblender-method}).

Traditionally, Bayesian inference for astronomical catalogs might use the conditional likelihood paired with Markov chain Monte Carlo (MCMC) sampling \citep{schneider2015hierarchical,portillo2017improved,feder2020multiband}.  Given the light source counts in an image and a parametric source model, one can run an MCMC chain over all sources and source parameters targeting the conditional likelihood $P(\Xv \vert \Zv)$ to output a joint posterior distribution of the sources' parameters. Although this fully Bayesian approach promises exact posterior samples upon convergence, MCMC sampling is less computationally tractable over larger parameter spaces and data volumes compared with a variational inference approach \citep{blei2017variational}.

Finally, the evidence, also known as the marginal likelihood and $P(\Xv)$, is the marginal probability distribution of the pixel values in a given image. $P(\Xv)$ is intractable for our case. We can rewrite this term as 
\begin{equation}
    P(\Xv) = \int P(\Zv) P(\Xv \vert \Zv) d\Zv,
\label{eq:evidence}
\end{equation}
where the integral is over all possible catalogs. Intuitively, the marginal likelihood is the probability of a given astronomical image without reference to a specific corresponding catalog. 
Hypothetically, it would be the probability of obtaining a given set of pixels (an image) when pointing the telescope at infinite random realizations of the sky. 
The pixels in $\Xv$ are correlated due to the effect of the PSF and the existence of extended sources, so the probability $P(\Xv)$ does not factorize.
This distribution reflects two sources of uncertainty: (1) the uncertainty over catalogs from the prior and (2) the noise in the image. 
This integral is intractable because the number of catalogs we need to consider to perform this integral is arbitrarily large for a given image $\Xv$. For example, a given set of pixel values in an image might be degenerate between any number of faint sources buried under noise, or any number of sources that sufficiently overlap for that resolution.


\section{Appendix B: Variational loss function derivation} \label{app:vi-derivation}
In this appendix, we derive the loss function used in BLISS corresponding to Equation~\ref{eq:vi-loss} in Section~\ref{sec:vi-in-bliss}.

We start with the FAVI loss function (Equation \ref{eq:initial-vi-loss}):
\begin{equation}
    L(\vphi) = \mathbb{E}_{\Xv \sim P(\Xv)} \KL{P(\Zv \vert \Xv)}{Q_{\vphi}(\Zv)},
\end{equation}
where $P(\mathcal{Z} \vert \mathcal{X})$ is the true posterior, $Q_{\vphi}(\mathcal{Z})$ is the variational distribution, $\mathcal{Z}$ are the light source parameters (Equation~\ref{eq:partial-catalog}), $\mathcal{X}$ is an astronomical image, and $\KL{\cdot}{\cdot}$ is the KL divergence.

The goal is to find the neural network weights $\vphi^{\star}$ that minimize this loss. We can formulate this mathematically as follows:
\begin{equation}
    \vphi^{\star} = \argmin_{\vphi} \mathbb{E}_{\mathcal{X} \sim P(\mathcal{X})} \KL{P}{Q_{\vphi}},
\end{equation}
where $P$ and $Q_{\vphi}$ are shorthand for $P(\Zv \vert \Xv)$ and $Q_{\vphi}(\Zv)$, respectively. We can now expand the KL divergence using the definition (Equation~\ref{eq:kl-divergence}) to obtain:
\begin{align}
    \vphi^{\star} &=\argmin_{\vphi} \mathbb{E}_{\mathcal{X} \sim P(\mathcal{X})}  \int P \log \left( \frac{P}{Q_{\vphi}} \right) \\
    &= \argmin_{\vphi} \mathbb{E}_{\mathcal{X} \sim P(\mathcal{X})}  \int P \log P - \mathbb{E}_{\mathcal{X} \sim P(\mathcal{X})} \int P \log Q_{\vphi}.
\end{align}
The first term can be ignored since it does not depend on $\vphi$ and thus will not affect optimization. This gives:
\begin{align}
    \vphi^{\star} &= \argmin_{\vphi} - \mathbb{E}_{\mathcal{X} \sim P(\mathcal{X})} \int P (\Zv \vert \Xv) \log Q_{\vphi} \\
    &= \argmin_{\vphi} -\mathbb{E}_{\mathcal{X} \sim P(\mathcal{X})} \mathbb{E}_{\mathcal{Z} \sim P (\mathcal{Z} \vert \mathcal{X})} \log Q_{\vphi}(\mathcal{Z}).
\end{align}
We can combine the two expectations to obtain: 
\begin{align}
    \vphi^{\star} &= \argmin_{\vphi} -\mathbb{E}_{\mathcal{X, Z} \sim P(\mathcal{X, Z})} \log Q_{\vphi}(\mathcal{Z}) \\
    &= \argmin_{\vphi} -\mathbb{E}_{\mathcal{X, Z} \sim P(\mathcal{X, Z})} \sum_{t=1}^{T} \log Q_{\vphi}(\Zv_{t}). \label{eq:klpq-loss}
\end{align}
Next, using the rules of conditional probability for the variational distribution with each light source parameter, we get:
\begin{equation}
    Q_{\vphi} (\Zv_{t}) = Q_{\vphi} (n_{t}) Q_{\vphi} (\ellv_{t} \vert n_{t}) Q_{\vphi}(b_{t} \vert n_{t}, \ellv_{t}). \label{eq:variational-distribution-factorizes-in-params}
\end{equation}
Finally, plugging this into Equation~\ref{eq:klpq-loss} we recover the loss used in BLISS for the detection and classification encoders (Section~\ref{sec:detection-classification-method}) and obtain:
\begin{align}
        L'(\vphi) &= -\mathbb{E}_{\mathcal{X, Z} \sim P(\mathcal{X, Z})} \sum_{t=1}^{T} \Big[ \log Q_{\vphi}(n_{t}) + \log Q_{\vphi}(\ellv_{t} \vert n_{t}) \nonumber \\ 
    &+ \log Q_{\vphi}(b_{t} \vert n_{t}, \ellv_{t}) \Big],
\end{align}
which corresponds to Equation~\ref{eq:vi-loss}. 

\section{Appendix C: Settings for SourceExtractor in Python} \label{app:sep-settings}

We use SExtractor in Python (SEP) v$1.4.1$ \citep{barbary2016sep}. The convolution kernel (FILTER) corresponds to a Gaussian profile in a $3\times3$ array of pixel values with a width of $2.0$ pixels ($0.4$ arcsec). 

The specific arguments used for detection through the \texttt{sep.extract} function are listed in Table~\ref{app:sep-settings}. 
We aim to match the SExtractor \citep{bertin1996sextractor} settings used in \cite{sanchez2021effects} as much as possible.

\begin{table}[h]
\centering
\begin{tabular}{|c | c|}
\hline
\textbf{Argument} & \textbf{Value} \\
\hline \hline
\texttt{thresh} & $1.5$ \\
\hline
\texttt{min\_area} & $5$ \\
\hline
\texttt{deblend\_nthresh} & $32$ \\
\hline
\texttt{deblend\_cont} & $0.005$ \\
\hline
\end{tabular}
\caption{
    Arguments passed into the \texttt{sep.extract} function, which roughly corresponds to classical SExtractor settings. See Appendix~\ref{app:sep-settings} for details.
}
\label{tab:sep-settings}
\end{table}

\section{Appendix D: Centroid prediction for pair blend} \label{app:pair-blend-centroids}

In this appendix, we show the quality of the centroid predictions of the detection encoder in the pair blend example of Section~\ref{sec:toy-blend-results}. 

\begin{figure*}[!hbtp]
    \centering
    \includegraphics[width=0.95 \textwidth]{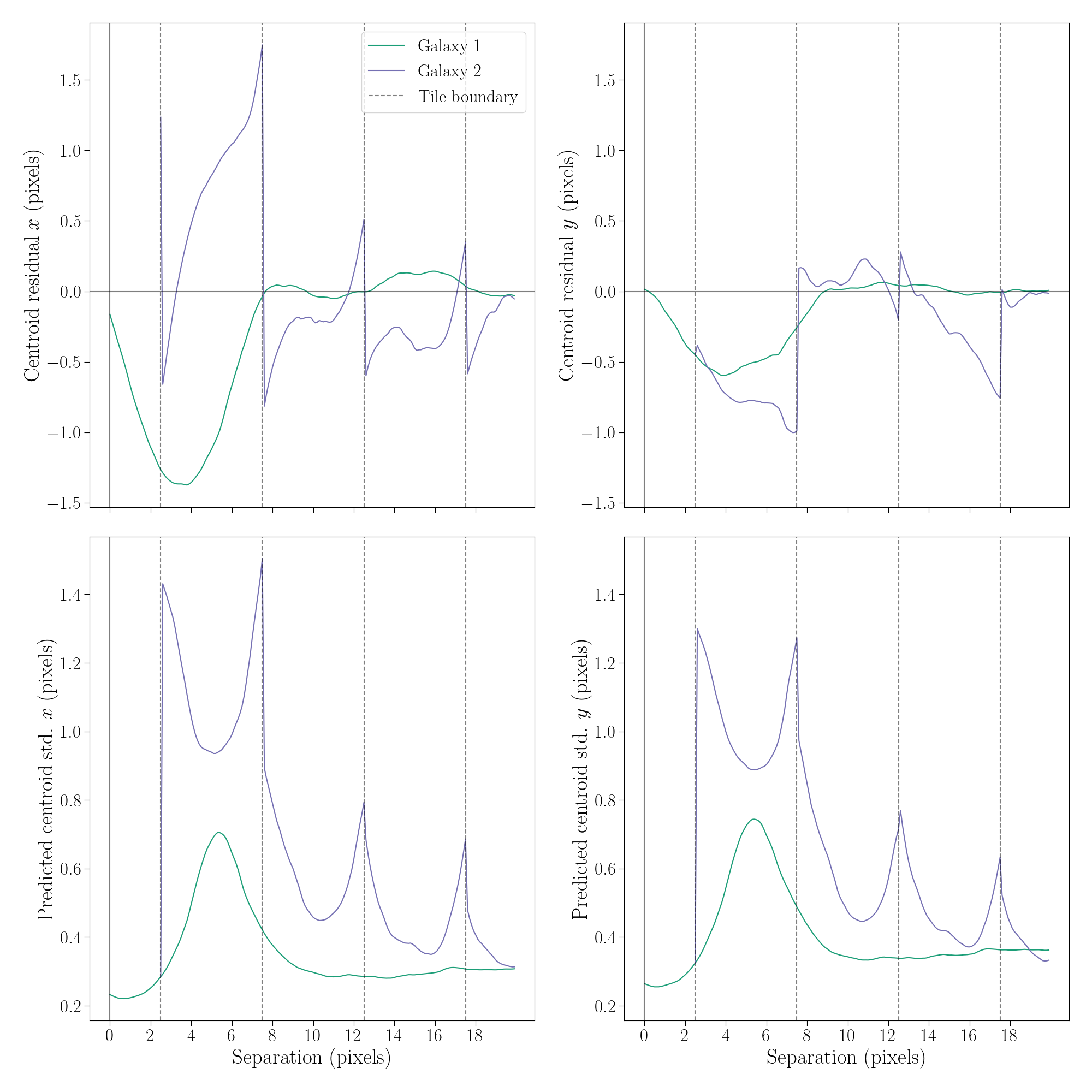}
    \caption{
        \textbf{Detection residuals and errors as a function of separation between galaxies.} In this plot, we show the centroid location residual, and BLISS predicted uncertainties in the $x$ and $y$ directions for the pair of galaxies in Section~\ref{sec:toy-blend-results}.
        Specifically, the top row shows the difference between the mean centroid predicted by the detection encoder for the tile containing each galaxy and the true centroid as a function of separation.
        The bottom row shows the predicted uncertainty on the centroid as a function of separation.
    } 
    \label{fig:toy-centroid-residuals}
\end{figure*}

In Figure~\ref{fig:toy-centroid-residuals}, top row, we see the residual between the true centroid and mean of the centroid posterior that BLISS predicts for the tile containing ``Galaxy 1'' (green) and ``Galaxy 2'' (purple). In the bottom row, we see the uncertainty in the centroid predicted by BLISS for the tile containing each source. Both the residuals and uncertainties are significantly noisy, but all seem to decrease with increased separation between the sources. 
We also see spikes in all the curves at separations corresponding to tile boundaries, which were also observed for the detection probabilities in Figure~\ref{fig:toy-detection-prob}.

\bibliography{references}
\bibliographystyle{aasjournal}

\end{document}